\documentclass[12pt]{scrartcl}
\usepackage{rotating} 
\usepackage{amsmath}
\usepackage{natbib}
\usepackage{graphicx}
\usepackage{geometry}
\usepackage{setspace}
\usepackage{tabularx}
\usepackage{multirow}
\usepackage{comment}
\usepackage{hyperref}

\title{Estimating Causal Effects with Observational Data: 
Guidelines for Agricultural and Applied Economists}

\author{
    Arne Henningsen\textsuperscript{1}, 
    Guy Low\textsuperscript{2}, 
    David Wuepper\textsuperscript{3},\\
    Tobias Dalhaus\textsuperscript{2}, 
    Hugo Storm\textsuperscript{3},
    Dagim Belay\textsuperscript{1},
    Stefan Hirsch\textsuperscript{4}}

\date{\small
    \textsuperscript{1}~Department of Food and Resource Economics, University of Copenhagen, Denmark;\\
    \textsuperscript{2}~Business Economics Group, Wageningen University \& Research, The Netherlands;\\
    \textsuperscript{3}~Institute for Food and Resource Economics, University of Bonn, Germany;\\
    \textsuperscript{4}~Department of Management in Agribusiness, University of Hohenheim, Germany\\[2ex]
    \normalsize August 4, 2025
}

\begin{document}

\maketitle

\thispagestyle{empty}

\vspace{-5ex}

\begin{spacing}{1}
\begin{itshape}
\small\noindent
This is a revised and extended version of the working paper:
Henningsen, A., G.~Low, D.~Wuepper, T.~Dalhaus, H.~Storm, D.~Belay, and S.~Hirsch (2024): 
Estimating Causal Effects with Observational Data: Guidelines for Agricultural and Applied Economists. 
IFRO Working Paper 2024/03, 
Department of Food and Resource Economics, University of Copenhagen,
available at
\url{https://econpapers.repec.org/RePEc:foi:wpaper:2024_03}.
\end{itshape}
\end{spacing}

\vspace{2ex}

\noindent
\textbf{Abstract}

\noindent
Most research questions in agricultural and applied economics are of
a causal nature, i.e., how one or more variables (e.g., policies, prices,
the weather) affect one or more other variables (e.g., income, crop yields,
pollution). 
Only some of these research questions can be studied experimentally.
Most empirical studies in agricultural and applied economics thus rely on observational data.
However, estimating causal effects with observational data requires
appropriate research designs and a transparent discussion of all identifying assumptions, together with empirical evidence to assess the probability that they hold.
This paper provides an overview of various approaches that are frequently used in
agricultural and applied economics to estimate causal effects with
observational data. It then provides advice and guidelines for
agricultural and applied economists who are intending to estimate causal effects
with observational data, e.g., how to assess and discuss the chosen
identification strategies in their publications.

\medskip\noindent
\textbf{Keywords}:  
causal inference, 
observational data, 
instrumental variables, 
difference in differences,
regression discontinuity,
synthetic controls

\medskip\noindent
\textbf{JEL codes}:
C21, 
C23, 
C24, 
C26, 
C51, 
C52 

\clearpage

\section{Introduction}

Today, around 50\% of empirical economics articles focus on causal
inference \citep{Imbens2024} and this number is even higher in applied economics.
However, a commonly observed problem in empirical research is that there is not always an obvious path to causal identification.  
Sometimes, one might only be able to approach causality without fully reaching it, e.g., it is possible to adjust for some confounders, but not all, or one manages to deal with reverse causality, but there is still the risk that systematic measurement error biases the estimate away from the true causal effect.
It is always important to discuss the main identifying assumptions of the chosen empirical approach and to assess the probability that they hold.
In practice, however, it is common that econometric estimates are causally interpreted without paying attention to the validity of the assumptions that allow this \citep{Gibson2019}. 
\citet{McKenzie2010} compared experimental and non-experimental methods in an empirical application
and found
that estimates from Ordinary Least Squares (OLS) regression, matching approaches, and difference-in-differences (DID) methods based on observational data overstate the effect of interest by 20--82\% compared to
an experimental benchmark. 
Sometimes, such biases can be revealed by combining causal inference approaches based on different identifying assumptions, such that if one set of assumptions does not hold, this is caught by the second approach.  

Thorough consideration of causality is of the utmost importance
when conducting empirical economics research \citep{Imbens2024}. 
The misinterpretation of statistical
associations as causal effects, together with insufficient robustness
and replicability of empirical analyses have motivated the ``credibility revolution'' in quantitative economics research and a call for higher
standards in statistical identification \citep{Angrist2010,Bellemare2012,Gibson2019}.%
\footnote{ 
More general factors related to the credibility of published results include publication bias, insufficient sample size, insufficient standardisation 
of variable definitions across studies,
and various types of statistical malpractice
such as p-value hacking
or exploratory research that is incorrectly framed as confirmatory research,
which is also known as Hypothesising After the Results are Known (HARKing).
All of these factors and practices can lead to biased and less replicable results and misleading conclusions. 
See \citet{Ioannidis2013} for a more detailed discussion of these drivers of incredibility.
}
While the ``credibility revolution'' has its origin
in labour economics, it has also reached agricultural
and applied economics, albeit with delay \citep{Bellemare2012}. 
Here, it can still frequently be
observed that empirical results that are used to test specific
hypotheses on the relationship between economic variables are
interpreted causally using terms such as ``effect'' or ``impact''
although the underlying research design and econometric framework are
not based on a valid identification strategy,
or at least not a sufficiently described and
motivated identification strategy. 
For example, some studies use OLS or
matching methods, which rely on a selection-on-observables assumption,
in a context with strong selection-on-unobservables. 
The use of these methods possibly moves
the estimates in the direction of the actual causal effect but often
not sufficiently far that the estimates can be causally interpreted. 
Other examples are studies that use a method based on instrumental variables (IVs), 
such as 2-stage least squares (2SLS) or endogenous switching regression, 
but do not sufficiently discuss or justify the validity of the IVs, 
or studies that use a difference-in-differences design but do not scrutinise the parallel trends assumption they are relying on.
The mere application of an econometric approach without sufficient verification
of the underlying assumptions is often falsely regarded as a
sufficient condition for the causal interpretation of the results.
Incorrect use of causal identification approaches may even make the
estimate worse and move it away from the actual causal effect. 
Examples are an erroneous null-finding because the parallel trends assumption for the
chosen DID estimator does not hold, or an exaggerated statistical
significance because the instrumental variable does not produce a strong
first stage.

Correct identification of causal effects is particularly relevant for agricultural and applied economics research because decisions by policy makers, international organisations (e.g., FAO, IFAD, World Bank), NGOs, and the private sector (e.g., agribusinesses, farmers) in the area of agriculture and food often have crucial effects on society, for example, through consequences for environmental sustainability, food safety, and food security \citep{Finger2023}.%
\footnote{
While recent research documents that the uptake of scientific evidence by policymakers may be selective \citep{Rao2025},
we are unaware of studies
that investigate the use of results from economics research by other decision makers.
However, given that governments and other stakeholders frequently spend money on independent research studies and that private businesses can benefit from reliable and unbiased scientific evidence,
we contend that a notable share of decision makers is genuinely committed to evidence-based decision making.
For example, in many resource-constrained settings, credible estimates of programme effectiveness can help guide more efficient allocation and support better-targeted interventions. 
Even where evidence uptake may be delayed or selective, credible (causal) impact evaluation solidifies the evidence base and can gradually shape policy discourse and decisions.
}
Hence, empirical agricultural economics papers
that aim to identify causal effects should include a clear description
and justification of the underlying ``identification strategy''. 
This refers to the identification of the exogenous variation in an endogenous
covariate or treatment variable of interest, i.e., the part of the
variation in this variable that is not related to unobserved confounders
\citep[e.g.,][]{Gibson2019,Lal2024}.
Only for this part of the
variation in the endogenous covariate or treatment variable,
it is possible to say that it \emph{affects} the dependent
variable \citep[e.g.,][]{Gibson2019}. 
Moreover, the limitations of
the identification strategy should be clearly outlined and possible
implications for the reliability of the results should be investigated.%
\footnote{ 
In addition, the external validity of the results should be outlined and discussed, 
e.g., whether the results that are based on a specific group of economic agents 
such as farmers or consumers in a specific region or country 
may also be valid for other groups of economic agents 
such as farmers or consumers in other regions or countries. 
However, the discussion of external validity is outside the scope of this paper,
and we only discuss internal validity.
} 
If a specific method is used to address the non-experimental
nature of the data, the added value compared to simpler
approaches such as OLS should be highlighted. 
If the added value cannot
be clearly shown, it may be preferable to stick with a simpler method and interpret the results as associations. 
Especially problematic are analyses in which an outcome is regressed on a set of explanatory variables and each coefficient is interpreted as reflecting the causal effect of the respective variable. 
This is usually inappropriate, as in most empirical applications, it is impossible to present a credible identification strategy for multiple explanatory variables and to avoid ``bad controls'' for estimating the causal effect of each of these explanatory variables.

The ``gold standard'' for internal validity is the use of randomised controlled trials (RCTs) \citep{Gibson2019} 
and numerous examples can be found in the agricultural and applied economics literature
\citep[e.g.,][]{Bulte2014,Wilebore2019}. 
However, RCTs also have important limitations \citep[see, e.g.,][]{Barrett2010}.
For example, most of the highly relevant research questions in agricultural and applied economics cannot
be answered with experiments because they would be problematic, impractical, or
infeasible for various reasons. 
For example, randomly assigning import tariffs,
randomly assigning different levels of education to future farmers at their birth,
increasing food prices in randomly selected regions, or
restricting food aid to specific regions while excluding others that are also in need \citep[][p.~36]{ALNAP2016} would be problematic for multiple reasons.%
\footnote{ 
Note that even if such experiments were feasible, 
it may be hard to prevent the non-treated group 
from becoming informed about the treatment of the intervention group 
\citep{ALNAP2016,Koppenberg2023}.
\citet{Deaton2010} and \citet{Bulte2020} provide overviews on the limitations of RCTs.
Furthermore, even when using experiments, only relationships with randomised variables can indicate causal effects,
while relationships with non-randomised variables (e.g., personal characteristics) usually cannot be interpreted as causal effects
\citep[see, e.g.,][as an example]{Nigus2024}.
Even in the relatively rare cases in which experimental methods can be
applied, their results often have important limitations. 
For example, RCTs are usually restricted to narrow cases, the results are regularly not directly generalisable, and there are often additional complications, such as non-compliance with the treatment and uncontrollable external influences. 
In addition, it is difficult to identify
the mechanisms behind the cause-effect interplay \citep{Quisumbing2020,Koppenberg2023,Todd2023}.
}
However, highly relevant research questions should not be neglected just because they
cannot be answered by applying experimental methods. Instead,
observational data must be used to answer these research questions
as thoroughly as possible.

This paper discusses various research designs and empirical methods that
are frequently used in agricultural and applied economics to estimate
causal effects with observational data. These discussions should help
researchers, analysts, and reviewers assess the suitability of these
empirical approaches in their specific analysis, choose the most
appropriate approach, justify their choice of approach, and
interpret their results appropriately. 
Therefore, we extend previous
literature that provides overviews \citep{Imbens2024} or guidelines on how to conduct
econometric identification methods using instrumental variables
\citep[e.g.,][]{Jiang2017,Young2022,Lal2024} for different disciplines,
and tailor our guidelines to research questions and commonly used
econometric approaches in agricultural and applied economics. 
We focus on the most common empirical research designs used in agricultural economics.
For focus and brevity, we omit approaches
that are less frequently used in agricultural economics so far,
e.g., the regression kink approach \citep{cattaneo2022regression}, 
bunching \citep{Dureti2025},
and the front-door criterion \citep{Bellemare2024}.  

The following section discusses the use of various methods that are
based on the `selection on observables' identification strategy such as
ordinary-least squares (OLS) and matching methods (e.g., propensity score matching). 
The third section explores methods that are based on
instrumental variables (or exclusion restrictions) such as 2SLS
regression and endogenous switching regression. 
The fourth section
discusses fixed-effects estimations and difference-in-differences approaches. 
The fifth section describes the synthetic control method, while the sixth section examines regression discontinuity designs. 
Finally, the seventh section concludes the paper and provides some general guidelines for agricultural and applied economics research.

\section{Selection on Observables}
\label{sec:selection-on-observables}

The selection-on-observables identification strategy is based on the
assumption that we observe and control for all variables that are
correlated with both the treatment and the error term. This implies that
there are no unobserved factors that are correlated with the treatment
and affect the outcome through pathways that are not blocked by control
variables. This assumption is also sometimes called conditional
independence assumption (CIA), conditional ignorability, or conditional
unconfoundedness.

Classical regression analyses (e.g., ordinary least squares (OLS),
logit, probit, tobit, or Poisson regression) can be affected by three
potential sources of statistical
endogeneity:%
\footnote{ 
In this paper, we focus on the endogeneity of explanatory variables. 
However, all other assumptions
that are required tor obtain unbiased and/or consistent estimates
should also be fulfilled and discussed
when presenting econometric analyses. 
For instance, the functional form used in the econometric analysis
should resemble the relationship between the explanatory variables
and the dependent variable in the population. 
Furthermore, the observations used for the estimation should be a random sample of the relevant population, 
while deviations from random sampling, 
e.g., non-proportional stratified random sampling, 
should be appropriately addressed in the econometric analysis. 
Furthermore, what the used data actually measure
and what the results really imply should also be correctly interpreted
\citep{Gibson2019}.
} 
(a)~omitted variables / unobserved heterogeneity; 
(b)~measurement error (any type of measurement
error in the explanatory variable or non-random measurement error in the
dependent variable), 
and (c)~reverse causality / simultaneity from which
it follows that the dependent variable also influences the explanatory
variable of interest. When discussing potential endogeneity in a
regression analysis, it is advisable to focus on each of the three
potential reasons separately \citep[see, e.g.,][]{BellemareNovak2017}.
Theoretically, all the explanatory variables must be uncorrelated
with the error term, while in practice the discussion of endogeneity
usually focuses on one or a few explanatory variables that are of
particular interest for the research question, e.g., treatment
variables. 
If a control variable is correlated with the error term, 
the bias of the estimated coefficient(s) of interest depends on the
relationship between this endogenous control variable and the
explanatory variable of interest, i.e., whether there is a direct
correlation or indirect relationship through other control variables
\citep[see][the latter provides an illustrative example with only one control variable]{Frolich2008, Bellemare2015}.%
\footnote{
Regarding the interpretation of the coefficients of covariates 
see \citet{Westreich2013}.
}

Whether a selection-on-observables identification strategy is feasible can be assessed, for example, by using Directed Acyclic Graphs (DAGs). 
DAGs are useful for at least two purposes. 
First, they are a useful tool to clearly communicate and discuss assumptions about relationships between variables. 
Second, by applying certain rules or algorithms to DAGs (either manually or through available software tools%
\footnote{
Several online and offline software tools for visualising and analysing DAGs exist. 
One of these tools is the open-source software DAGitty
(\url{https://www.dagitty.net/}).
}),
sets of suitable control variables can be determined \citep{Morgan2014, Pearl2018}.%
\footnote{
It is important to note
that a DAG indicates whether a causal effect
is \emph{non-parametrically} identified,
i.e., the identification does not rely on parametric assumptions,
e.g., about the functional form of the modelled relationships
or the distribution of the error term.
Even when using parametric empirical methods,
in most cases it is desirable to identify causal effects non-parametrically
so that approximately reliable results are obtained
also if parametric assumptions are not 100\% fulfilled.
}
This also includes the identification of variables
that should \emph{not} be used as control variables, i.e., variables on the
causal path from the treatment variable to the outcome variable (``bad
controls''). 
DAGs were originally developed in computer science \citep{Pearl2018}, but are increasingly being used in economics \citep{Imbens2020, Hunermund2025}. 
However, it is important to emphasise that a DAG should not be considered as the only ``true'' and universally valid presentation of the real world, but rather as a tool to communicate the underlying assumptions of an empirical analysis. 

Some studies aim to address unobserved heterogeneity by using a control
variable that indicates the marginal utility of joining or leaving the
`treatment' \citep{Verhofstadt2014,BellemareNovak2017,Ruml2021,Aihounton2024}. Theoretically,
this approach seems promising, but in practice it can be
problematic because the control variable is usually observed after the
decision to participate in the treatment has been made and, thus, it can be influenced by
the treatment itself, which can introduce endogeneity \citep{Aihounton2024}.

Some empirical researchers try to address endogeneity by using lagged
values instead of concurrent values of explanatory variables. \citet{Bellemare2017} show theoretically that using lagged values of explanatory
variables addresses endogeneity only under the untestable assumption of
``no dynamics among unobservables''. Their Monte Carlo simulation shows
that using lagged values of explanatory variables can result in
substantially biased estimates and incorrect inference even if there
are only low levels of dynamics among unobservables \citep{Bellemare2017}. Providing convincing arguments that there are no dynamics in any
unobservable variables seems to be very difficult or impossible for most
empirical studies.

Using matching methods such as propensity score matching
(PSM)%
\footnote{ 
\citet{King2019} point out that
``propensity scores should not be used for matching'' 
and that other matching methods are more suitable than PSM.
} 
or inverse probability weighting for estimating causal effects with observational data is
basically based on the same identifying assumptions as regression
methods \citep[e.g.,][]{Angrist2009,Blattman2010,Mullally2018}. Therefore, the same discussion as for the use of
regression methods is required. The same applies to the augmented
inverse propensity weighted (AIPW) estimator which is `doubly-robust' as
it basically requires the same identification strategy as an OLS
regression \citep[e.g.,][equation 1]{Kurz2022}.

There are methods for assessing the sensitivity of the results to
unobserved heterogeneity \citep[e.g.,][]{Oster2019,Cinelli2020,Diegert2023},
which have been used often in recent applied economics research. 
However, these methods are, in general, based on bold assumptions, and it is difficult or impossible to assess 
whether these assumptions are fulfilled in a specific empirical application. 
However, when applying a selection-on-observables identification strategy, 
these methods can contribute to assessing the suitability of the identification strategy
if their assumptions are discussed appropriately and their results are
interpreted carefully.

Classical regression methods usually rely on strict assumptions about
the functional form of the relationship between treatment variables,
control variables and the dependent variable. These restrictive
assumptions can be relaxed by using nonparametric regression methods,
most of the available matching methods, or machine learning approaches.
While machine learning methods have rapidly advanced and are being
increasingly used in agricultural and applied economics, it is important
to point out that most machine learning methods are unsuitable when they are used directly to estimate causal effects even if all variables that are correlated with
both the outcome and the treatment variable are observed. This is
because machine learning methods are generally designed for prediction
and not the direct estimation of causal relationships. 
For example, machine learning approaches for variable selection (such as Lasso) 
select the subset of covariates that optimises out-of-sample prediction performance, but this selection likely introduces omitted-variable biases as it drops highly correlated control variables, 
including covariates that are correlated with both the outcome and the treatment variable. 

However, machine learning methods can be used within established econometrics
frameworks for causal identification such as under the
selection-on-observables assumption or for IV estimation 
(see Section~\ref{sec:IV} and Appendix Section~\ref{sec:IV-extended-methods}). 
These methods are then called ``causal machine learning.''
Despite this name, it should be clear that these methods are not new concepts for causal
identification but rather extensions of the established econometrics
frameworks of causal identification in which specific parts are replaced
by machine learning methods. 
Hence, they come with the same identification assumptions that apply to ``classical''
econometric approaches and, thus, the same requirements to carefully
consider and motivate an appropriate identification strategy. The
basic idea of causal machine learning is to leverage the predictive
capabilities of machine learning methods and their flexibility to
approximate potentially complex relationships within these frameworks
\citep{Storm2020,Baylis2021}. 
For example, under the
selection-on-observables assumption, causal machine learning methods can be used to relax restrictive functional form assumptions such as in the case of Double/Debiased Machine Learning (DML) \citep{Chernozhukov2018}, which
assumes that the outcome model is a separable additive function, but
that treatment effects, the influence of controls on
outcomes, and the treatment assignment are unknown nonlinear functions.
The approach allows the use of any machine learning algorithm to
approximate these nonlinear functions and to derive average treatment
effects. 

The ``Causal Forests'' method \citep{Wager2018}, which is a special case
of Generalised Random Forests \citep{Athey2019}, extends the DML approach
allowing the estimation of heterogeneous treatment effects, i.e., treatment
effects that depend on observed characteristics (conditional average
treatment effects, CATE). From an applied perspective, a crucial
advantage is that treatment heterogeneity is estimated in a transparent
and data-driven way and thus avoids the need to predefine and potentially
cherry pick treatment groups. In agricultural economics, Causal Forests
have already been applied in various contexts to study treatment heterogeneity
\citep[e.g.,][]{Deines2019,Stetter2022,Deines2023,Schulz2024},
while \citet{Brignoli2024} conduct simulation studies to compare the performance of classical econometric methods, Causal Forests, and other machine-learning methods in the estimation of (heterogenous) treatment effects with typical cross-sectional farm-level data.

In summary, when relying on a selection-on-observables identification strategy,
we suggest doing the following
(in addition to following the general suggestions
that we provide in Section~\ref{sec:conclusion}):
\begin{itemize}
\item Clearly state the assumptions that the chosen method and model specification require for obtaining unbiased and/or consistent estimates.
\item Use a DAG to find a suitable model specification (e.g., which control variables to include and which not to include) and to discuss the credibility of the chosen identification strategy.
\item Separately discuss the three potential sources of statistical endogeneity: (a)~omitted variables / unobserved heterogeneity, (b)~measurement error, and (c)~reverse causality / simultaneity.
\item Discuss the potential statistical endogeneity not only of the explanatory variable of interest but also of the control variables.
\item Consider using methods for assessing the sensitivity of the results to unobserved heterogeneity.
\item Consider using methods that do not rely on strict parametric assumptions.
\end{itemize}

\section{Instrumental-Variable Methods}
\label{sec:IV}

Instrumental-variable methods are often used in cases in which selection on
observables cannot be justified \citep{Lal2024}. We define
`instrumental-variable (IV) methods' in a broad sense. 
While this section focusses on the use of IVs in linear IV and 2-stage least squares (2SLS)
regression (which is identical to IV-regression if the number of IVs%
\footnote{
In this paper, we use the narrow definition of IVs,
i.e., we only consider the variables that are used to explain the endogenous regressor
but that are not used to explain the outcome variable as IVs,
while the broad definition of IVs
additionally includes the variables that are used to explain the outcome variable
because these variables are also used to explain the endogenous regressor.
}
is equal to the number of endogenous regressors),
these discussions and the practical advice given in this section also apply to other estimators that rely on IVs,
including machine-learning IV methods
(see Appendix Section~\ref{sec:IV-extended-methods}). 
A brief overview of special types of instrumental variables is presented
in Appendix Section~\ref{sec:IV-special-types}.

The assumptions required by IV approaches are sophisticated and
difficult to test empirically \citep{Lal2024}. However, this does not
imply that we want to discourage their use, rather our
aim is to provide some suggestions and tools on how to implement
credible IV-based identification strategies in empirical research.
This is important as invalid instruments can exacerbate the problem, so that
the bias in the 2SLS estimator even exceeds the OLS endogeneity bias
\citep{Lal2024}. 
By construction, IV estimates are less precise than
OLS estimates. 
For example, \citet{Lal2024} analyse 70~IV designs and show that 2SLS estimates have, on average, six times higher standard errors than OLS estimates, 
although this decreases with instrument strength.%
\footnote{  
See Figure~3 of \citet{Lal2024}. 
\citet{Lal2024} also point out that this makes 2SLS estimations more susceptible to p-value hacking and publication bias.
}

Using an instrumental-variable approach to estimate a causal effect is
possible if one has at least as many instrumental variables as
endogenous regressors.
These instrumental variables must fulfil the following two
criteria: (a)~they must be ``relevant'', i.e., strongly related to the
endogenous regressors and (b)~they must be statistically ``exogenous'',
i.e., not related to the error term (exclusion restriction).

The first criterion can be empirically investigated with tests for weak instruments.
Traditionally, an instrumental variable was considered to
be relevant (i.e., not weak) if an F~test of its relevance in the
first-stage regression had a test statistic of 10 or higher \citep{Staiger1997}. 
However, more recent research indicates that a test
statistic of 10 is insufficient in most empirical applications. For
instance, \citet{Keane2024} show that OLS estimates are often closer
to the `true' causal effects than 2SLS estimates if the F-statistic of
the first stage is below~20. 
They also demonstrate that in cases in which there is only one instrument, 
the evaluation of instrument strength should be based on an
F-statistic that exceeds~50. 
Moreover, estimation results (e.g., t-tests) are often unreliable even in cases in which there are much higher values for the F-statistic \citep[e.g.,][]{Lee2022,Keane2023,Keane2024}.
In addition, \citet{Lal2024} show that first-stage F-statistics are
frequently overestimated if the test is not robust towards heteroskedasticity, clustering and autocorrelation, 
which implies that IVs in such cases may incorrectly be treated as relevant.

The exclusion restriction implies that the exogenous (excluded) instrument
influences the dependent variable only via its effect on the endogenous
explanatory variable and it is not correlated with the error term.
If the endogenous explanatory variable is continuous,
the exogeneity of the instrumental variables cannot be
empirically investigated without further assumptions
\citep{Pearl1995a,Pearl1995b,Gunsilius2021}.%
\footnote{
Some empirical researchers \citep[e.g.,][]{Acemoglu2001}
aim to test the exogeneity of instrumental variables
by estimating the outcome equation with both the endogenous regressor
and the instrumental variable (and of course all relevant control variables).
If the instrumental variable affects the dependent variable only through the
endogenous regressor, the coefficient of the instrumental variable
in this auxiliary regression should be close to zero.
However, if the endogenous explanatory variable is indeed endogenous,
the coefficient of this variable and the coefficient of the instrumental variable
are not jointly identified \citep{Conley2012}.
Hence, this auxiliary regression does not provide useful information.
}
For instance, if there are more potential instrumental variables
than endogenous regressors are available,
it is possible to apply the Sargan-Hansen test / Sargan's \emph{J} test / Hansen's \emph{J} test for overidentifying restrictions.
If, based on theoretical considerations, it is certain that there are at least
as many \emph{exogenous} instrumental variables as there are endogenous regressors,
the test indicates (under some assumptions, e.g., correct model specification)
whether the additional instrumental variables,
i.e., those that are not certain to be exogenous, 
are indeed exogenous. 
However, without clear theoretical justifications that ascertain the exogeneity
of at least as many IVs as there are exogenous regressors,
the test is basically uninformative.

In contrast, if the endogenous explanatory variable is discrete,
the exogeneity of the instrumental variables can be tested.
\citet{Pearl1995a,Pearl1995b} derives testable inequalities,
which have been extended by \citet{Kedagni2020}.
The intuition behind these inequalities is
that for observations with the same value of the endogenous explanatory variable,
the value of the dependent variable should not depend on the value of the instrument.
While these inequalities have been very rarely used in empirical research,
some researchers tested the exogeneity of instrumental variables
by combining this intuition with parametric assumptions.
For instance, \citet{DiFalco2011} apply a `falsification test'
for IV-regression with a binary endogenous explanatory variable
that indicates treatment status.
In this `falsification test', the regression model is re-estimated
with untreated observations only and with the endogenous regressor
replaced by the instrumental variable. 
A valid instrument should not be statistically significant in this regression.
The same procedure could be repeated for the treated observations.
In case of a categorical endogenous explanatory variable,
the procedure could be conducted for observations in each category of this variable.

In addition, it is helpful to think of placebo estimates that can be used to test specific violations of the exclusion restriction. 
For instance, the instrumental variable might affect the treatment via a specific mechanism
that only matters for some observations (e.g., specific locations, farmers, or crops) but not for others. 
In this case, a useful placebo test would be to obtain reduced-form estimates of the correlation between the outcome and the instrumental variable for a (sub)sample of observations, where the outcome and the instrumental variable should be uncorrelated. 
If the main concern is that the instrumental variable might
affect the outcome through a specific pathway other than the endogenous
regressor, and this potential other pathway is measurable, one can
directly test this potential violation of the exclusion restriction by
regressing this pathway on the instrumental variable. 
For example, if an
instrumental variable is supposed to affect the farmers' access to credit but is assumed not to affect their access to insurance, one can regress
farmers' access to insurance on the instrumental variable. 

One weakness of all the tests mentioned above is that they can never `prove' that an IV is exogenous because they all have the null hypothesis that the instrumental variables are exogenous
and not rejecting the null hypothesis does not necessarily mean that it is true,
particularly if the test has low statistical power,
e.g., caused by a small number of observations,
multicollinearity, or a large error variance. 
Hence, it is always necessary to strongly motivate the exogeneity of instrumental variables based on solid theoretical argumentation \citep[e.g.,][]{Lal2024} and critically discuss the assumption of statistical exogeneity
for each instrumental variable used, e.g., by debating potential
(unobserved) variables that may be related to both the treatment
variable and the outcome variable. 
This is very important as, for example, \citet{McKenzie2010} show that using instruments for which the exclusion restriction is potentially violated may lead to the overestimation of the effect of up to 82\% compared to the effect found from an experimental benchmark study. 
This is more than the overestimation that occurs when simply applying OLS (35\%), matching (20\%) or DID (22\%), 
which implies that a badly identified 2SLS estimation only makes things worse. As a general rule, the less specific the effects of the chosen instrumental variable, the less likely the exclusion restriction is valid (see, e.g., \citet{Mellon2024} for a discussion of rainfall as an instrument).

In the case of a weak instrument or a violation of the exclusion
restrictions, an IV estimation can lead to greater bias than an OLS
regression \citep{Lal2024}. In such cases, it is advisable to apply
non-causal estimators, interpret the results as associations, and
draw conclusions with due caution. Here we refer, e.g., to \citet{Groher2020}
and \citet{Aihounton2024} for examples of correlational
wording. \citet{Lal2024} note that 2SLS estimates are in many cases
much larger than standard OLS estimates although the aim of the IV
estimation is usually to tackle a positive omitted variable bias of OLS.
It is, therefore, advisable to also discuss the direction of the bias that
the IV estimation is intended to address and assess the extent to
which the IV approach was able to address this bias \citep[for examples, see, e.g.,][]{Basu2018,Hirsch2023}.

For estimating 2SLS, modern statistical software offers various
packages. It is advisable to use these rather than manually estimating
2SLS by first estimating the first-stage OLS and then manually inserting
the predicted values into a separately estimated second-stage OLS
regression. A common mistake when using the `manual' procedure is failing to include the same control variables in both stages, which
results in inconsistent 2SLS estimates \citep{Angrist2009}.
Furthermore, the `manual' procedure results in incorrect OLS standard
errors in the second stage. However, unless the instruments are very
strong, even the standard errors obtained by software packages for 2SLS
estimations do not correctly reflect the uncertainty of 2SLS estimates
and, thus, they need to be further adjusted \citep{Lee2022,Lal2024}. 
For example, \citet{Lal2024} analyse 70~IV designs and report that the estimated standard errors of 2SLS estimates systematically underestimate the uncertainties of these estimates.

For the interpretation of results, it is important to note 
that 2SLS estimates indicate average treatment effects (ATE) only under restrictive assumptions (e.g., that the treatment effect is homogeneous across all subjects with the same values of the control variables) \citep[e.g.,][]{Heckman1997,Aronow2013}.%
\footnote{ 
\citet{Aronow2013} suggest a method 
that requires either homogeneity of the treatment effect or homogeneity of compliance (i.e., instruments have the same effect on the treatment assignment across all observations).
} 
However, these assumptions are unlikely to be fulfilled in many empirical analyses.
Under less restrictive assumptions (e.g., monotonicity of the effect of
the instrumental variable on the endogenous explanatory variable), 2SLS
estimates indicate local average treatment effects (LATE), which
indicates the effect of the part of the variation in the endogenous
explanatory variable that is caused by variation in the instrumental
variable \citep[e.g.,][]{Imbens1994}. For instance, in the case of a
binary instrumental variable and a binary endogenous explanatory
variable, the LATE indicates the average treatment effect on those
subjects that `comply' with the instrumental variable, while the effects
on the `always takers' and the `never takers' remain unidentified
and in most cases it is unknown who the `compliers' actually are. 
While the LATE may provide relevant information in some empirical analyses,
in others it might not identify the effect we are interested in \citep{Angrist2009,Aronow2013}.

Although the above discussions refer to linear IV and 2SLS regression,
they are largely transferable to a large number of other methods that rely on instrumental variables or exclusion restrictions
such as endogenous switching regression models
or methods for non-continuous dependent or endogenous explanatory variables
(see Appendix Section~\ref{sec:IV-extended-methods} for details).
It is important to note 
that additional pitfalls exist
when using instrumental variables in regression models with non-linear terms (e.g., quadratic, interaction terms) 
and/or in non-linear regression models
(e.g., probit, logit)
(see Appendix Section~\ref{sec:IV-extended-methods} for details).

While the availability of a valid instrument is a crucial requirement for
obtaining unbiased treatment estimates using any IV approach, 
it is also crucial to consider the functional form assumption that underlies
the employed methods. 
For instance, \citet{Okui2012} show that 2SLS
regression may result in substantially biased estimates of the treatment
effect if the functional relationship between the control variables and
the outcome variable is incorrectly specified. Interestingly, in applied
settings, much of the discussion seems to focus on the validity of the
instrument, while often the strong functional form assumptions seem to
be more readily accepted and less critically discussed. However,
depending on the degree of heterogeneity or nonlinearity, they may be
equally critical \citep{Okui2012}.

Existing nonparametric versions of IV estimators relax these functional form assumptions and require only
that the outcome is the sum of an (unknown) nonlinear function of a treatment variable and observed covariates
(that are uncorrelated with unobserved confounders)
and an additive error term
that may be correlated with the treatment variable \citep{Newey2003}. 
However, early nonparametric approaches based on basis functions/splines or
kernel methods struggle with a larger number of covariates or
instruments and large sample sizes.
Building on these early nonparametric estimators, 
an active field of research at the intersection of machine learning and econometrics has developed extensions
that leverage the predictive capabilities of modern machine learning methods to improve nonparametric IV estimators.

Although these new machine learning-based IV approaches offer some interesting extensions of existing approaches, 
it is important to emphasise that they do not change the requirement of having a solid identification strategy and valid instruments.%
\footnote{
Appendix Section~\ref{sec:IV-extended-methods} provides a more detailed discussion of machine learning IV methods.
}

Generally, the promise of IV estimation is that it can estimate unbiased effects despite
unobserved confounders. However, any IV approach comes at the cost of a substantial reduction in
the statistical power of the estimation. This
is particularly relevant to consider when estimating heterogeneous
treatment effects 
(given that estimating not just one value but infinitely many or
a function of values is a substantially more complex task). Hence,
applying IV methods with the aim of identifying treatment heterogeneity
typically requires large datasets.

If various assessments indicate that an IV-based method should be considered, we suggest performing the following checks
that comprise a combination of theory-based considerations and
suitable statistical tests \citep[e.g.,][]{Lal2024} 
(in addition to following the general suggestions
that we provide in Section~\ref{sec:conclusion}):
\begin{itemize}
\item If an explanatory variable is incorrectly treated as endogenous,
estimates based on IV regression (e.g., 2SLS) are less efficient
than estimates based on corresponding selection-on-observables regression methods (e.g., OLS). 
Therefore, it is important to consider and discuss,
based on theoretical argumentation,
whether a potentially endogenous explanatory variable should indeed be instrumented.
In all cases, it is advisable to provide and compare the results for both the IV regression and the OLS estimation.
\end{itemize}

When using an IV regression method,
it is important to assess the strength of the instruments based on the following criteria:
\begin{itemize}
\item Always report complete first-stage results including all model diagnostics.
\item Only use IV-based methods when the IV(s) are sufficiently strong,
i.e., there is a sufficiently high correlation between the endogenous explanatory variable and the IV(s) after controlling for exogenous control variables.
\item Assess the strength of the IV(s) by applying an F-test to the first stage of the IV estimation
that tests the statistical significance of the IVs
(i.e., tests a specification with the exogenous control variables but not the IV(s) as explanatory variables against the complete first-stage regression with the IV(s)).
\item If the F-statistic of the statistical significance of the IVs in the first stage is below 20, consider presenting OLS estimates instead of 2SLS estimates
as OLS estimates are often closer to the `true' causal effects than are 2SLS estimates.
In the case of a single instrument, the F~statistic should exceed 50 \citep{Keane2024}.
\item If the first-stage F~statistic is below 100, standard errors may need to be adjusted as described by \citet{Lee2022} or \citet{Keane2024}.
\item In the case of heteroskedasticity, clustering, or autocorrelation of the error term in the first stage, 
it is important to conduct an F-test that is robust to
these conditions, as a standard F-test overestimates the F~statistic \citep{Lal2024}. 
See, for example, the Cragg-Donald F~statistic \citep{Cragg1993}
or the Kleibergen-Paap statistic \citep{Kleibergen2006} 
and the guidance on these statistics provided,
e.g., in \citet{Bazzi2013} or \citet{Windmeijer2024}.
\end{itemize}
We refer to previous parts of this section and the literature \citep[e.g.,][section~2.2.1]{Lal2024} 
for a more in-depth discussion of the options to investigate the strength of IVs.

If the instruments are sufficiently strong (so that the use of IV
regression is not abandoned), it is important to assess the
appropriateness of the exclusion restriction / independence assumption.
We suggest doing the following:
\begin{itemize}
\item Use strong theoretical considerations to rule out any direct effect of the IVs on the dependent variable or any relationship with omitted factors (error term),
see, e.g., \citet{Mellon2024}, who discusses the use of weather as an instrument.

\item If the endogenous explanatory variable(s) is (are) discrete,
use statistical tests to test the exogeneity of the IV(s),
e.g., the tests suggested by \citet{DiFalco2011} or \citet{Kedagni2020}.

\item If possible, use placebo tests to assess the exclusion restriction(s)
(see above).

\item In the rare situations when the model is overidentified (i.e., the number of IVs is larger than the number of endogenous explanatory variables)
and there are clear theoretical justifications for the exogeneity
of at least as many IVs as there are exogenous regressors,
use a Sargan-Hansen test / Sargan's \emph{J} test / Hansen's \emph{J} test
to test the exogeneity of the additional instrumental variables,
i.e., those that are not certain to be exogenous.
However, in most cases, one cannot be certain that there are at least as many exogenous instruments as there are endogenous regressors,
and, thus, the result of this test gives no practical guidance.
Furthermore, it is important to note that this test relies on a correct model specification and does not investigate instrument relevance.

\item Be aware of the limitations of statistical tests for the exogeneity of IVs,
particularly that not rejecting the null hypothesis of exogeneity
does not mean that the IVs are exogenous,
particularly if the test has low statistical power.
\end{itemize}
For further discussion on how to assess the exclusion restriction,
we refer to previous parts of this section and the literature
\citep[e.g.,][section~2.2.2]{Lal2024}.

If the exclusion restriction / independence assumption is considered to
be appropriate, it is important to carefully assess and interpret the
second-stage results and:
\begin{itemize}
\item Provide OLS estimates for comparison.
\item Discuss whether 2SLS was able to address the bias of OLS estimates, which involves a discussion of the direction of the bias and the extent to which a 2SLS regression can attenuate this bias \citep[see, e.g.,][]{Basu2018}.
\item Interpret the results as LATE unless there is credible evidence
that the chosen method and empirical specification provide an estimate of the ATE.
\item Use the tF test \citep{Lee2022} or the Anderson-Rubin (AR) test \citep{Keane2024} instead of standard t-tests.
\end{itemize}

\section{Fixed Effects and Difference in Differences}
\label{sec:fe_did}

Fixed effects are a useful tool to control for unobserved confounders
that are constant at the fixed-effect level. 
For example, when using individual-fixed effects in a study with panel data, 
which in agricultural economics papers are often farm-fixed effects, 
one can control for all time-invariant unobserved heterogeneity at the individual (farm) level. 
The unobserved heterogeneity may be differences in management skills, local climatic and soil conditions, infrastructure, or the remoteness of the area.
Consequently, models with individual-fixed effects cannot quantify the effects of time-invariant factors such as
proximity to a city \citep{Wooldridge2010}. 
Similarly, fixed effects can be set and combined at every level that reasonably groups the data. 
For instance, year-fixed effects control for all unobserved heterogeneity that affects all units in a given year in the same way, such as market conditions, the introduction of a certain policy, etc.. 
Mathematically, fixed effects are equal to a joint demeaning of the dependent variable and the independent variables, which is also called \textit{within transformation}. 
For farm-fixed effects, this implies subtracting the farm average from each observation. 
This transforms, for instance, farm profits  into deviations from the average profit of the respective farm in the observed time period \citep{Mundlak1961}. 
Fixed effects may be helpful for controlling
for many unobserved factors, and they may also be combined with other methods such as IV or DID. 
However, there are only a few examples of cases in which fixed effects are
sufficient to fully establish causality in a model \citep{Blanc2017}. 
One example is weather shock impact models
that regress a measure of agricultural performance
such as yields or productivity
on a random and exogenous weather shock \citep{Blanc2017}.
Remaining caveats of fixed-effect models are related to reverse causality and time-variant confounders,
which may still introduce simultaneity and omitted-variable biases.%
\footnote{
While fixed effects help to control for biases arising from unobserved time-invariant confounders, 
common issues in fixed-effect applications are temporal and spatial correlation, clustering, and heteroscedasticity in the error term.
The standard approach to dealing with this is to obtain standard errors that are robust to these deviations from independently and identically distributed errors \citep[see, e.g., ][for an example]{Low2025}.
}

Taking a closer look at the above examples of time-invariant factors, 
climatic conditions, soil quality, and infrastructure
may be reasonably considered time-invariant in the short run but they may change over longer time horizons. 
Therefore, \citet{Millimet2023} 
follow \citet{Mundlak1961,Mundlak1978} and argue that such a potential bias
may be ignored in shorter panels due to negligible changes in these variables over time. 
However, in increasingly long panels, a trade-off arises between efficiency gains derived from more observations and potential biases and inconsistency resulting from not truly time-invariant factors
accumulating to considerable unobserved confounders over time. 
\citet{Millimet2023} highlight alternative estimators such as the First-Difference~(FD), Twice First-Differenced~(TFD) or Interactive Fixed Effects~(IFE) estimator, and suggest Rolling FD~(RFD), Rolling TFD~(RTFD), and Rolling FE~(RFE) estimators, 
which can either be used as alternatives to Fixed-Effect (FE) estimators
or at least to explore the sensitivity of FE estimates to these alternative estimators. 

Table~\ref{tab:FE} summarises panel data estimators that address unobserved heterogeneity.
When using panel data sets with two time periods~($T=2$),
`rolling' estimators cannot be used, while (individual) FE estimates are equal to FD estimates without intercept and two-way fixed effects~(TWFE) estimates are equal to FD estimates with intercept.
In case of more than two time periods ($T>2$) and mostly time-invariant unobserved heterogeneity,
FE, TWFE, and FD estimators are recommended.
The FD estimator is preferable if there is strong positive temporal autocorrelation in the untransformed error term
because the first differencing removes this strong temporal autocorrelation,
while FE and TWFE estimators are preferable if there is no or only little temporal autocorrelation in the untransformed error term
because in this case first differencing would introduce strong negative temporal autocorrelation.

The robustness of estimators that rely on the assumption of time-invariant confounders can be assessed by comparing their estimates to those of estimators that are more robust to time-varying unobserved heterogeneity.
If discrepancies arise, \citet{Millimet2023} recommend additionally reporting the results of alternative estimators. 
In the very long run ({$T \to \infty$}), most unobserved heterogeneity would change over time, making FE and TWFE unreliable and less relevant \citep{Bellemare2025}. In agricultural economics, many outcomes, such as crop or milk yield and farm profits, exhibit low autocorrelation, making FE or TWFE estimators appropriate. 
In these settings, using panel data with higher frequencies is often more beneficial than merely extending the time dimension of the panel \citep{Millimet2023}. 
For example, \citet{BelayOlsen2025} leverage monthly data to implement TWFE and IFE models in their analysis of milk yield.

\begin{table}[htbp]
\caption{Panel Data Estimators for Unobserved Heterogeneity (\(N \gg T\))}
\label{tab:FE}
\renewcommand{\arraystretch}{1.5} 
\scriptsize
\begin{tabular}{|p{16mm}|>{\raggedright\arraybackslash}p{24mm}|>{\raggedright\arraybackslash}p{43mm}|>{\raggedright\arraybackslash}p{30mm}|>{\raggedright\arraybackslash}p{24mm}|}
\hline
\textbf{Time \(T\)} & 
\textbf{Unit Unobserved Heterogeneity} & 
\textbf{Recommended Estimator} & 
\textbf{Implementation in STATA} & 
\textbf{Implementation in R} \\
\hline
\(T = 2\) & 
(Mostly) time-invariant  & 
FE = RFE = FD without intercept;\newline TWFE = FD with intercept & 
\texttt{xtreg}, \texttt{reghdfe}, \texttt{gen} + \texttt{by} (manual) & 
\texttt{fixest}, \texttt{plm} \\
\hline
\multirow{5}{*}{$2 < T \ll \infty$} & 
(Mostly) time-invariant & 
FE and TWFE (if no or little autocorrelation),
FD (if strong positive autocorrelation) & 
\texttt{xtreg}, \texttt{reghdfe}, \texttt{gen} + \texttt{by} (manual) & 
\texttt{fixest}, \texttt{plm} \\
\cline{2-5}
 & 
Gradual time-varying & 
FD, RFD, RFE & 
\texttt{rolling}, \texttt{rangestat}, \texttt{reghdfe} in loop & 
\texttt{plm}, \texttt{fixest} + \texttt{rollapply} \\
\cline{2-5}
 & 
Rapid time-varying & 
TFD, RTFD, RFE & 
\texttt{rolling}, \texttt{rangestat}, \texttt{reghdfe} in loop & 
\texttt{rollapply}, \texttt{fixest} in loop \\
\cline{2-5}
 & 
Linear time-varying & 
FE and TWFE (with unit-specific trends), 
IFE for flexible form & 
\texttt{xtreg}, \texttt{reghdfe} & 
\texttt{fixest}, \texttt{phtt} \\
\cline{2-5}
 & 
Non-linear time-varying & 
IFE, RTFD & 
\texttt{regife}, \texttt{reghdfe} & 
\texttt{phtt}, \texttt{rollapply}, nonparametric rolling \\
\hline
\multirow{2}{*}{$T \to \infty$} & 
Strictly time-invariant (rare in practice) & 
FE, TWFE, FD & 
\texttt{xtreg}, \texttt{reghdfe} & 
\texttt{fixest}, \texttt{plm} \\
\cline{2-5}
 & 
Mostly time-varying  (incl.\ formerly invariant) & 
RTFD, TFD, RFE, IFE & 
\texttt{rolling}, \texttt{rangestat}, \texttt{reghdfe} in loop & 
\texttt{rollapply}, \texttt{phtt}, \texttt{fixest} in loop \\
\hline
\end{tabular}
\textbf{Note:} 
FE = Fixed Effects; 
TWFE = Two-Way Fixed Effects; 
FD = First Differences; 
TFD = Twice FD; 
RFD = Rolling FD; 
RFE = Rolling FE; 
RTFD = Rolling TFD;
IFE = Interactive Fixed Effects. 
This requires panel data with observations from units $i \in \{1, \ldots, N\}$ over time periods $t \in \{1, \ldots, T\}$, typically with $N \gg T$. All these recommendations assume homogenous treatment effects.
\end{table}

An alternative approach to estimating causal effects with panel data is the difference-in-differences (DID) research design.%
\footnote{
It is important to note
that DID is a \emph{research design},
while FE, TWFE, FD, etc.\ are \emph{estimation methods}.
Depending on the data structure and assumptions about the data generating process,
different estimators are suitable for DID research designs.
}
In classic ($2 \times 2$) DID estimations,
there are two groups and two time periods.
There is a pre-treatment period, 
when no units are treated; 
and there is a post-treatment period, 
when some units are treated (the treated group) 
and others (the control group) remain untreated. 
By using the control group as the
counterfactual in the post-treatment period,
it is possible to calculate the average difference between the observed effects of a treatment and
the counterfactual: the ``average treatment effects on the treated'' (ATT).

The underlying identifying assumption in DID is the parallel-trends assumption, 
which reasons that the treated units would have followed the same parallel trends as the untreated control units
had the treated units gone from the pre-treatment period to the post-treatment period in the
absence of treatment.%
\footnote{
In certain cases, a simple DID design 
may not yield reliable causal inference. 
For instance, if a policy targets farmers younger than 40~years in a specific state, 
comparing this group of farmers to either farmers aged 40--49~years in the same state
or to farmers younger than 40~years in other states
may lead to biased estimates
because it does not account for age-related or state-specific trends, respectively. 
To address this, a triple-DID estimator uses differences in three dimensions (state, age group, and time) 
to isolate the causal effect of the policy change.
The triple DID estimator, 
which can also be calculated as the difference between two DID estimators, 
may only require one parallel trend assumption
as long as the bias is the same in both estimators,
in which case the bias cancels out when differenced \citep{Olden2022}.
}
If this assumption is satisfied, then the control
units can provide the counterfactual for the treated group in the post-treatment period. 
However, the parallel-trends assumption is purely hypothetical by definition 
since it is impossible to be certain 
that the trends of the treated units and the untreated control units would
have followed parallel paths in the post-treatment period. 
When a data set includes multiple pre-treatment periods,
one can verify
that the pre-treatment trends of the two groups are parallel, 
though one should be cautious when inferring ``true causality'' 
as parallel trends in the pre-treatment periods may not necessarily imply parallel trends
between the last pre-treatment period and the post-treatment period in the hypothetical situation in which the treatment group is not treated. 

Multiple applications of DID in agricultural and food economics settings exist.
For instance, in production economics, \citet{Belay2020} estimate the effect of information disclosure on antibiotic use and market survival among pig farms, 
while \citet{BelayAyalew2020} examine the impact of reference market price disclosure on smallholders' crop choice. 
Similarly, \citet{Belay2022} evaluate the impact of limiting antibiotic use on the economic performance of pig farms. 
In consumption economics, 
\citet{Fan2022} estimate the impact of the introduction of a sugar tax on candy purchases and
\citet{Hoy2020} estimate the impact of GMO labelling on consumer choices. 
Other studies, such as \citet{Wuepper2022} apply DID design to evaluate alternative agri-environmental schemes. 

The basic $2 \times 2$ DID set-up can be extended to situations with multiple time periods. 
In DID settings with multiple time periods, 
a key question is how treatment effects evolve with exposure duration,
i.e., do they increase, decrease, or remain stable over time? 
Investigating these dynamics is often the primary reason for using event-study (ES) regressions in DID designs \citep{Callaway2021, Miller2023}.
The DID estimations with multiple periods can also be extended to scenarios,
in which different units of the treatment group receive the treatment at different times, 
which is known as heterogeneous treatment timing. 
Under conditions in which the size of the treatment
effect is (a)~constant over time and (b)~independent of the time period of the treatment, a standard two-way fixed effects estimator offers a
reliable estimation for inferring treatment effect causality
\citep{Rothetal2023}.

However, under heterogeneous treatment timing and treatment effect heterogeneity,  
the two-way fixed effect estimator may results in a biased estimate of the average treatment effect on the treated 
and, thus, causally interpreting the regression coefficient becomes problematic 
even if the parallel-trends assumption holds \citep{deChaisemartin2020,GoodmanBacon2021,Athey2022}. 
For instance, this may be the staggered%
\footnote{
Staggered treatment is a setting
where different units  adopt/implement the treatment at different times with no reversal to the unit's treatment status,
i.e., if a unit is treated once, it remains always treated \citep{Callaway2021}.
} 
adoption of an agricultural policy whose effect is time-varying,
i.e., the magnitude of the effect depends on the time 
when a farm faced the treatment (e.g., policy) for the first time,
the number of years that the farm has already faced the treatment
(e.g., due to adjustments, learning, and/or accumulating effects over time),
and/or the specific year (e.g., on the weather or market conditions in the year).
By making so-called ``forbidden comparisons''
between groups that received the treatment at earlier and later times,
standard DID methods may give negative estimates of the average treatment effect on the treated 
even when the `true' effect is, in fact, positive, which is known as the
negative weights problem \citep{GoodmanBacon2021,deChaisemartin2023a,Borusyak2024b}. 
Recent developments in DID have identified solutions to this issue. Studies by \citet{Callaway2021}, 
\citet{Sun2021}, \citet{Wooldridge2021}, \citet{deChaisemartin2023b}, and \citet{Borusyak2024b}
have overcome the negative weights problem by restricting the types of
comparisons that can be made and ensuring that appropriate
counterfactuals are used to causally infer effects under
heterogeneous treatment timing and treatment effect heterogeneity under
various conditions of the parallel-trends assumption. 
For example, one may condition the parallel-trends assumption on additional
covariates, such as weather or growing conditions, or on anticipatory
behaviour such as in the event of an upcoming policy change
\citep{Callaway2021}. 

Researchers could choose from alternative DID estimators summarised in Table~\ref{tab:DID}, 
depending on the treatment design, 
data structure, number of groups,
causal estimand of interest (e.g., overall/static ATT or event-study/dynamic ATT),
choice of baseline period (in event studies),
control group definition (in staggered designs), nature of parallel trends violation,
computational speed, and other criteria. 
The table includes several recent heterogeneity-robust DID estimators for staggered treatment designs. 
For example, in the context of gradual policy rollout, 
one can select either the never-treated group or the not-yet-treated group as controls \citep{Callaway2021,deChaisemartin2023b}.
A researcher can opt for estimators that construct counterfactuals using imputation methods based on efficient and fast linear estimation \citep{Borusyak2024b}, 
two-stage difference in differences \citep{Gardner2024},
or non-linear DID models such as exponential, logit, or probit models \citep{Wooldridge2021}. 
Moreover, heterogeneity-robust DID designs have also been developed for continuous (i.e., non-binary and
non-discrete) treatments \citep{Callaway2024a, Callaway2024b, deChaisemartin2024,deChaisemartin2024b, deChaisemartin2025}
as well as for multiple (i.e., reversible and re-treatable) \citep{deChaisemartin2024} and several treatments \citep{deChaisemartin2023a}. 
In the case of multiple treatments (sometimes also called treatment-on-and-off scenario),
it is important to distinguish between ``no carryover'' and ``(arbitrary) carryover.''
In the ``no-carryover'' case, 
only the current treatment status affects outcomes
with no lasting impact from past treatment \citep{deChaisemartin2024}. 
In contrast, ``(arbitrary) carryover'' means 
that the treatment history influences outcomes,
making it resemble the staggered treatment scenario. 
In this case, ``intent-to-treat'' effects can be estimated 
by defining treatment as ``has ever been treated'' 
in a staggered treatment fashion, 
thereby ensuring that the treatment status is absorbing and accounts for any potential carryover effects \citep{Sun2021,Liu2024}. 
In many cases, the effect of having previously received the treatment is of interest
as it reflects the long-term impact of the treatment, 
even if the treatment itself is temporary. 
For instance, \citet{Deryugina2017} studies the fiscal cost for counties hit by hurricanes. 
Although hurricanes are transitory, their long-term impact persists, 
so she models the year of the first hurricane to capture these effects. 
She then adopts what \citet{deChaisemartin2023b} refer as a ``binarize and staggerize'' approach, 
i.e., by replacing the hurricane status (on/off) with a binary indicator of having been previously hit by a hurricane, 
the treatment becomes absorbing, allowing the use of staggered adoption designs \citep{Sun2021, deChaisemartin2023b}.

It is important to note that the estimation methods recommended for various DID model scenarios in Table~\ref{tab:DID}, 
along with their implementation in Stata and R, are based on the assumption 
that the parallel-trends assumption holds unconditionally (i.e., without covariates). 
From all methods listed in Table~\ref{tab:DID},
the method suggested by \citet{Callaway2021} is the most suitable for cases where the parallel-trends assumption holds
only after conditioning on covariates.
This method is applicable for treatments that are both binary and staggered. 

When violations of the parallel trends assumption arise from long-run discrepancies in outcome trends between groups,
estimators such as the one suggested by \citet{Borusyak2024b},
which leverage the full set of pre-treatment periods to construct counterfactuals, 
can be particularly effective, especially compared to methods that rely solely on the last pre-treatment period as a baseline.
However, if the violation stems from a known anticipation effect, 
approaches such as those suggested by \citet{Callaway2021} and \citet{Sun2021} can be adapted to use the anticipation period as a baseline to produce reliable estimates \citep{deChaisemartin2023b}.

Moreover, DID in an event study framework (including recent generalised DID estimators) offers plots that visually present both dynamic treatment
effects and pretreatment trends,
allowing the evaluation and testing of parallel trends before treatment \citep[e.g.,][]{Taylor2022,Li2024}. 
However, it is important to note that failure to detect a non-parallel pre-treatment trend does not necessarily imply its absence, as conventional event study
tests for parallel pre-treatment trends often lack power and therefore fail to detect non-parallel pre-treatment trends \citep{ Freyaldenhoven2019,Freyaldenhoven2021, Roth2022}. 
Researchers should assess the statistical power of these tests using tools such as the R package ``pretrends'' \citep{RothPretends} for nonlinear trends,
and consider alternative visualisation tools 
such as the ``xtevent'' package in Stata \citep{Freyaldenhoven2025} 
or the ``eventstudyr'' package in R \citep{Freyaldenhoven2023} 
or magnitude-based pre-treatment trend evaluation \citep{Bilinski2020}. 
If the (unconditional) trends during and after the treatment
cannot be considered to be parallel
(e.g., if pre-treatment trends do not seem to be parallel),
\citet{Freyaldenhoven2019} recommend using a 2SLS framework (available in ``xtevent''  or ``eventstudyr'') with one or more covariates that are affected by the confounding (non-parallel) trends but are not related to the treatment. 
\citet{Rambachan2023} propose confidence sets that are robust to violations of the parallel-trends assumption,
which can be obtained using the ``HonestDiD'' package \citep{RambachanHonestDiD} in R or Stata, as applied by \citet{Wuepper2022}.
Regardless of the approach, 
using economic knowledge to analyse potential parallel trend violations strengthens causal inferences over relying solely on the statistical significance of tests
of parallel pre-treatment trends \citep{Roth2022}.
Furthermore, it is important to emphasise
that the parallel-trend assumption cannot be tested,
as even perfect parallel pre-treatment trends do not guarantee
that the trends during and after the treatment period would also be parallel.%
\footnote{
Although parallel pre-treatment trends are neither necessary nor sufficient for obtaining unbiased estimates,
it is highly recommended to test for parallel pre-treatment trends
because if there are parallel pre-trends, 
it is more likely that the parallel trends assumption is fulfilled
and if pre-treatment trends are not parallel, it is less likely that the parallel trends assumption is fulfilled. 
}

An interesting extension to study staggered treatment problems is the
matrix completion approach for causal panel data models, which allows the combination of
two-way fixed-effects with synthetic controls in a data-driven
manner \citep{Athey2021}. 
In an agricultural context, this approach is
particularly appealing as it naturally deals with unbalanced panel
data sets \citep{Martinsson2024}. 
Similarly, \citet{Arkhangelsky2021} developed the Synthetic Difference-in-Differences (SDID) method, which combines elements of the synthetic control approach (discussed below) with DID. 
While SDID requires a relatively longer pre-treatment period to construct credible counterfactuals, 
it does more than merely testing for parallel pre-treatment trends using past outcomes \citep{Roth2022}; 
it leverages the pre-treatment information to compute weights that ensure pre-treatment parallel trends hold by construction.

Another relevant impact estimator related to DID design is the Changes-in-Changes (CiC) estimator \citep{Athey2006}, 
which serves as an alternative to DID by
focusing on the Quantile Treatment Effect on the Treated (QTT) rather than on ATT. 
The QTT approach helps policymakers understand
how the benefits and/or costs of a treatment are distributed across subgroups,
particularly when decisions depend on distributional effects. 
For example, \citet{Mayr2023} apply CiC to estimate heterogeneous impacts of voluntary climate agreements in the UK on business electricity consumption and employment.

When using fixed-effect-based or DID-based methods,
we suggest doing the following
(in addition to the general suggestions
that we provide in Section~\ref{sec:conclusion}):
\begin{itemize}
\item Provide reasoning based on economic theory on unobserved confounders
that potentially bias estimates and that can be addressed by the use of fixed effects.
\item Provide reasoning on the time invariance of potential unobserved confounders with respect to the covered time horizon when using individual-fixed effects. 
\item Select an appropriate estimator to account for unobserved heterogeneity in panel data, and justify your choice with compelling arguments (see, e.g., Table~\ref{tab:FE}).
\item When using FE or TWFE estimators (e.g., in case of low temporal autocorrelation), increasing data frequency is more beneficial than lengthening the panel duration.
\item Adjust standard errors to make them robust to heteroscedasticity, clustering, and spatial and temporal autocorrelation (if necessary).
\item Choose a suitable DID method and substantiate the choice of method by providing convincing arguments (see, e.g., Table~\ref{tab:DID}).
\item Evaluate if pre-treatment trends are parallel by creating parallel-trend plots in static DID analyses and event-study plots in dynamic DID settings.
\item Empirically investigate the extent to which pre-treatment trends are parallel in DID settings. This investigation should include supplementing event-study plots with diagnostic tests that assess the statistical power of tests for pre-treatment parallel trends.
\item Consider using methods such as those suggested by \citet{Abadie2005}, \citet{Santanna2020}, and \citet{Callaway2021} 
in DID settings where the parallel-trends assumption only holds when conditioning on covariates. 
However, \citet{Freyaldenhoven2019} emphasise
that this conditioning approach may often be inadequate in real-world economic applications 
because it requires the conditioning covariate to be a perfect proxy for the confounding trend---an assumption that may not always hold. 
To address this, \citet{Freyaldenhoven2019} propose a generalised 2SLS framework for an event study 
that allows conditioning on covariates 
that are not necessarily perfect proxies for the confounding trend.
\item Provide reasoning based on economic theory on post-treatment parallel trends in DID settings.
\item In DID with staggered treatment, consider using the Bacon Decomposition to explicitly diagnose and interpret static TWFE estimates as a weighted average of all possible pairwise $2 \times 2$ DID comparisons \citep{GoodmanBacon2021}.
This decomposition can be conducted with or without time-varying covariates, and implemented using the \textit{bacondecomp} package in Stata \citep{GoodmanBacon2022} or R \citep{FlackBacondecomp}. 
Alternatively, one can use the \textit{twowayfeweights} package in Stata \citep{deChaisemartin2024a} or  the \textit{TwoWayFEWeights} package in R \citep{CicciaTwoWayFEWeights}. 

\item Consider supplementing your DID estimates using falsification or placebo tests on outcomes arguably unrelated to the treatment/intervention. 
\end{itemize}

\begin{sidewaystable}[htbp]
\caption{Difference-in-Differences Methods}
\label{tab:DID}
\fontsize{7.5}{9} \selectfont
\begin{tabular}{|>{\raggedright\arraybackslash}p{12mm}|>{\raggedright\arraybackslash}p{9.5mm}|>{\raggedright\arraybackslash}p{18mm}|>{\raggedright\arraybackslash}p{25mm}|>{\raggedright\arraybackslash}p{17mm}|>{\raggedright\arraybackslash}p{25mm}|>{\raggedright\arraybackslash}p{35mm}|>{\raggedright\arraybackslash}p{35mm}|>{\raggedright\arraybackslash}p{35mm}|}
\hline
\bfseries  Time periods &
\multicolumn{2}{p{30mm}|}{\bfseries Treatment design} &
\centering\bfseries TE / Estimand &
\multicolumn{2}{p{32mm}|}{\centering\bfseries Specific design} &
\bfseries Recom\-mended estimator &
\bfseries Implementation in STATA &
\bfseries Implementation in R \\
\hline
Two &
Bi\-nary &
\multirow{2}{10mm}{Single treated group} &
\multirow{2}{20mm}{Static (ATT)} &
\multicolumn{2}{c|} { } &
(Static) TWFE, AA, SZ &
reghdfe, xtreg, absdid, drdid &
plm, fixest, DRDID \\
\cline{2-2} \cline{7-9}
 &
Conti\-nuous &
 &
 &
\multicolumn{2}{c|} { } &
CGS &
 &
 contdid \\
\hline
Multiple (Event Study (ES)) &
Bi\-nary &
Single treated group &
Static (Average of ES coefficients: ``Overall'' ATT) &
\multicolumn{2}{c|} { } &
(Static) TWFE, AA, SZ, AAHIW,  Methods in Dynamic/ES ATT &
reghdfe, xtreg, absdid, drdid, sdid, Packages in Dynamic/ES ATT  &
plm, fixest, DRDID, synthdid, Packages in Dynamic/ES ATT \\
\cline{4-9}
 &
 &
 &
Dynamic/ES ATT &
Baseline &
Average of all pre-treat\-ment periods &
BJS, W21, GT, AAHIW  &
did\_imputation, did2s, xthdidregress,
jwdid, wooldid, sdid\_event &
didimputation, did2s, etwfe, synthdid \\
\cline{6-9}
  &
  &
  &
  &
  &
Last pre-treat\-ment period  &
(Dynamic/ES) TWFE, CS, DH, SA  &
reghdfe, eventdd, xtevent, eventstudyinteract,
csdid,
did\_multiplegt\_dyn &
plm, fixest,
did,
\mbox{DIDmultiplegtDYN} \\
\cline{3-9}
 &
 &
Stag\-gered  &
Dynamic/ES ATT &
Baseline &
Last pre-treat\-ment period  &
CS, DH, SA  &
eventstudyinteract,
csdid,
did\_multiplegt\_dyn &
fixest,
did,
\mbox{DIDmultiplegtDYN} \\
\cline{6-9}
 &
 &
 &
 &
 &
Average of all pre-treat\-ment periods &
BJS, GT, W21, W23, AAHIW  &
did\_imputation, did2s, xthdidregress,
jwdid, wooldid, sdid\_event &
didimputation, did2s, etwfe, synthdid \\
\cline{5-9}
 &
 &
 &
 &
Controls &
Not-yet-treated group &
CS, DH  &
csdid, did\_multiplegt\_dyn    &
did, \mbox{DIDmultiplegtDYN}  \\
\cline{6-9}
 &
 &
 &
 &
 &
Last-treated or never-treated group &
CS, SA, AAHIW &
csdid, eventstudyinteract, sdid\_event &
did, fixest, synthdid \\ 
\cline{6-9}
 &
 &
 &
 &
 &
Imputation (regression-based) &
BJS, GT, W21, W23 &
did\_imputation,
did2s, xthdidregress, jwdid,
wooldid &
didimputation,
did2s, etwfe  \\
\cline{6-9}
 &
 &
 &
 &
 &
Imputation (synthetic control-based) &
AAHIW &
sdid\_event &
synthdid \\
\cline{5-9}
 &
 &
 &
 &
\multicolumn{2}{p{32mm}|}{Fast estimation} &
BJS, GT &
did\_imputation, did2s &
didimputation, did2s \\
\cline{5-9}
 &
 &
 &
 &
\multicolumn{2}{p{32mm}|}{Non-linear estimations} &
W23 &
jwdid, wooldid &
etwfe \\
\cline{5-9}
 &
 &
 &
 &
\multicolumn{2}{p{35mm}|}{(Quasi-)random assignment of treatment} &
RS, AI &
staggered &
staggered \\
\cline{3-9}
&
&
\multirow{2}{15mm}{Multiple (Treatment on-and-off)}  &
Static (``Overall'') ATT &
\multicolumn{2}{p{32mm}|}{No carryover} &
DH &
did\_multiplegt\_stat, did\_multiplegt\_dyn &
\mbox{DIDmultiplegtDYN}, \mbox{didmultiplegtstat}  \\
\cline{4-9}
  &
  &
  &
Dynamic/ES ATT &
\multicolumn{2}{p{32mm}|}{(Arbitrary) carryover } &
Methods in Staggered design, LWX  &
Packages in Staggered design, fect &
Packages in Staggered design, fect\\
\cline{2-9}
 &
 Conti\-nuous &
 Single treated group &
Static (``Overall'') ATT&
Stay\-ers &
Yes &
CGS, DHPSV &
did\_multiplegt\_stat &
contdid, \mbox{didmultiplegtstat} \\
\cline{6-9} 
 &
 &
 &
 &
 &
No &
CGS, DHV &
did\_had &
contdid, \mbox{did\_had} \\
\cline{4-9}
 &
 &
 &
 Dynamic/ES ATT &
Baseline &
Last pre-treat\-ment period  &
 CGS, DH &
did\_multiplegt\_dyn &
contdid, \mbox{DIDmultiplegtDYN} \\
\cline{3-9}
 &
 &
 Stag\-gered &
 Dynamic/ES ATT &
 Controls &
 Not-yet treated group &
CGS, DH &
did\_multiplegt\_dyn &
contdid, \mbox{DIDmultiplegtDYN} \\
\cline{5-9}
 &
 &
 &
 &
 Baseline &
 Last pre-treat\-ment period &
CGS, DH &
did\_multiplegt\_dyn &
contdid, \mbox{DIDmultiplegtDYN} \\
\hline
\end{tabular}
\fontsize{7}{8} \selectfont
\textbf{Notes:} 
AA: \citet{Abadie2005};
AAHIW: \citet{Arkhangelsky2021};
AI: \citet{Athey2022};
BJS: \citet{Borusyak2024b};
CGS: \citet{Callaway2024a};
CS: \citet{Callaway2021}; 
DH: \citet{deChaisemartin2024};
DHPSV: \citet{deChaisemartin2025};
DHV: \citet{deChaisemartin2024b};
GT: \citet{Gardner2024};
LWX: \citet{Liu2024};
RS: \citet{Roth2023};
SA: \citet{Sun2021}; 
SZ: \citet{Santanna2020};
TWFE: two-way fixed effects;
W21: \citet{Wooldridge2021};
W23: \citet{Wooldridge2023}.
All these recommendations are explicitly made under the assumption 
that the unconditional parallel trends assumption is fulfilled (without covariates).  
However, some of these methods are also suitable in DID settings, 
in which the parallel-trends assumption only holds when conditioning on covariates (e.g., AA, CS, and SZ).
\end{sidewaystable}

\section{Synthetic Control Method}
\label{sec:synthetic_control}

The synthetic control method (SCM) was introduced by \citet{Abadie2003}
and later formalised by \citet{Abadie2010,Abadie2015}. 
According to \citet{Athey2017}, 
``the synthetic control approach [...] is arguably
the most important innovation in the policy evaluation literature
in the last 15 years.'' 
However, despite the increasing availability of long panel datasets,
this method has not yet been widely applied in agricultural economics. 
The few examples for the use of this method in agricultural economics include, 
e.g., \citet{Grogger2017}, \citet{Mohan2017}, \citet{Opatrny2020} and \citet{Kim2023}.

SCM is basically a combination of DID and matching. 
Synthetic control units are selected as the weighted average of all potential comparison units based on how closely they resemble the treated unit(s) in the pre-treatment periods \citep{Abadie2021}. 
SCM is particularly valuable when the parallel trends assumption required for DID does not hold, provided that sufficiently long pre-treatment panel data are available. 
It is especially well-suited for evaluating the impact of interventions affecting a single or small number of large units such as cities, regions, or countries, 
making it a useful tool in agricultural and applied economics, 
where national or state-level agricultural, food, and environmental policies can be assessed by constructing a synthetic control group of comparable nations or states.
For example, \citet{Grogger2017} estimates the impact of the soda tax implemented in Mexico in 2014 on soda prices by comparing them to those of other untaxed non-substitute goods, creating a synthetic control group. 
Alternatively, researchers can construct the synthetic control using soda price data from other countries not subject to the tax, offering another way to estimate the causal effect of the policy. 

Furthermore, a balanced panel data set must be available
that includes the treated unit(s) and
a reasonably large number of potential comparison units (``donor pool''), 
while it includes a reasonably large number of pre-treatment periods
and at least one post-treatment period.%
\footnote{
Some Generalised SCM methods can also be applied to unbalanced panel data,
e.g., the method implemented in the R~package `gsynth' \citep{XuGsynth}
that adds the capability to use unbalanced panel data
to the method suggested by \citet{Xu2017}.
}
Although the synthetic control method usually cannot give
unbiased estimates of the treatment effect, 
\citet{Abadie2010} show that---under certain assumptions---the bias is bounded
and approaches zero with an increasing number of pre-treatment periods.
Hence, it is important to have a sufficiently large number of pre-treatment units.

One of the most basic assumptions of standard synthetic control methods is
that the data generating process corresponds to a `factor model'
\citep[equation~1]{Abadie2010},
which assumes, e.g., 
that unobserved differences between units are constant over time and
that the effects of observed and unobserved differences between units on the outcome
are identical across all units (but these effects can change over time). 
Thus, empirical applications must clearly discuss the appropriateness of these assumptions, 
e.g., if the treatment could potentially affect the effects
of observed and unobserved variables on the outcome
so that these effects differ between the treatment unit and the control unit
in the post-treatment period. 
Furthermore, in order to avoid overfitting,
the number of potential comparison units should not be too large,
which can be achieved by restricting potential comparison units to those
that are sufficiently similar to the treatment unit \citep{Abadie2015}.  

When using SCM, we suggest doing the following
(in addition to the general suggestions
provided in Section~\ref{sec:conclusion}):
\begin{itemize}
\item Make sure that there is a sufficiently long pre-treatment period.
\item Ensure that there is a sufficiently large but not too large number of comparison units. 
\item Visualise the SCM estimation results using graphs.
\item Present the contributions of each unit to the synthetic control. 
\item Conduct inference using the permutation method \citep{Abadie2015}.
\item To support the internal validity of causality using SCM, researchers are advised to conduct validity tests such as leaving out units of the donor pool (with non-zero weights), placebo tests, using fake treatment dates, and other outcomes not related to the treatment.
\item When appropriate and beneficial for reliability, consider combining SCM with DID, using SDID \citep{Arkhangelsky2021}.
\item Note that (a)~inference in SCM is limited to the data used to construct the synthetic control, (b)~SCM does not allow predictions or inferences outside the range of the observed data, and (c)~extrapolation or generalisation outside the supporting data and context is invalid. 
\end{itemize}

\section{Regression Discontinuity and Difference-in-\allowbreak{}Discontinuity Designs}

Regression Discontinuity Designs (RDDs) and Difference-in-Discontinuity Designs (DiDDs) can be set up in multiple ways
\citep[as discussed below and in][in more detail]{Wuepper2023}
but they all share a particular mechanism for identifying causal effects:
If treatment assignment is triggered by a clearly-defined threshold in a continuously distributed variable,%
\footnote{
Under certain conditions, 
it is also possible to apply the RDD framework
if the running variable is discrete
(e.g., food safety inspection score based on restaurant hygiene inspections).
However, the empirical analysis must take into account the discreteness of the running variable.
Details are available, e.g., in \citet{Kolesar2018}
and \citet{cattaneo2022regression},
while software packages such as
``rdhonest'' for Stata  \citep{armstrong2023} 
or ``RDHonest'' for R \citep{Kolesar2025}
can be used.
}
then---given a few falsifiable assumptions---discontinuity in the outcome 
right at this threshold quantifies the treatment effect
\citep{Thistlewaite1960, Imbens2008}.
Intuitively, this works especially well with arbitrarily set thresholds
because this minimises the risk that, besides the treatment assignment,
something else ``jumps'' exactly at the threshold. Another important
condition is that observations (usually people) cannot choose which
side of the threshold they are on (e.g., if it is well known that a subsidy
is available to farms below a certain size, farmers whose farms are just above the
threshold may be able to take measures that ensure that their farms fall just below, which might
make the treatment endogenous).

\begin{figure}[htbp]
\centering
\includegraphics[width=0.99\textwidth]{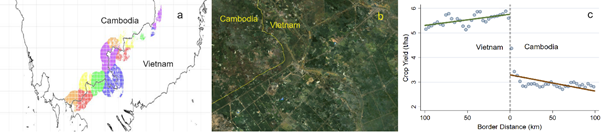}
\caption{(\textbf{a})~The border between Cambodia and Vietnam
separates an otherwise comparable agricultural area into two countries.
Colours distinguish different border segments. (\textbf{b})~Satellite
data can be used to obtain a methodologically unified, high-resolution
crop yield measure. (\textbf{c})~An important step: Before the actual
RDD is estimated, the data should be plotted, so that it is possible to visually inspect whether
the discontinuity that is to be estimated is visible. It is usually
helpful to aggregate the data points in small bins and fit regression
lines separately on both sides of the threshold. The actual RDD
estimates the size of the discontinuity at the threshold.\\
Sources: \citet{Wuepper2023a}~(a+c), Google Earth~(b)}
\label{fig:rdd}
\end{figure}

The fundamental requirement for Regression Discontinuity Designs (RDD)
is the existence of a continuously distributed variable that has a
threshold which triggers treatment
assignment.%
\footnote{
The threshold does not have to deterministically trigger the treatment 
as it does in the standard model. 
If the threshold only changes the probability of treatment, 
one moves from the sharp RDD to the fuzzy RDD, 
which involves estimating an instrumental variable regression
such as 2SLS with the threshold as the instrument.
} 
For instance, public extension services
may only visit farms within an arbitrarily defined maximum
distance-to-branch \citep{Pan2018}, and governments might target
villages with an anti-poverty programme if they are above an arbitrarily
defined poverty threshold \citep{AlixGarcia2013}. Also, geographical
borders can be used such as historical borders within a country \citep{Noack2022}, 
or national borders dividing countries
\citep{Wuepper2020b,Wuepper2020c}. 
When geographic borders are being used, 
the most general treatment one can define is 
``belonging to one side of the border or the other.''
For example, one might ask how much agricultural or environmental outcomes are simply the result of an area belonging to one country and not another
(see, e.g., Figure~\ref{fig:rdd} or \citealp{Wuepper2020b}).
When the border triggers mainly one specific mechanism, 
one might also be able to focus more narrowly on this mechanism directly.
For example, \citet{Noack2022} use the historical border between East and West Germany to identify the effect of agricultural structures (small-scale vs.\ large-scale farming) on bird diversity, 
and \citet{gupta2024} use Indian state borders to identify the negative impact of language barriers on the effectiveness of agricultural extension services.
Sometimes the treatment is introduced spatially with a clear boundary, 
e.g., in the case of protected areas \citep{neal2024} or World Heritage sites \citep{rodriguez2025}. 
In this case, the effect of
``belonging to one side and not the other'' is a narrow treatment itself.

The most intuitive way to understand how a national border
can be used to identify the effect of an area belonging to one country
but not another is provided in Figure~\ref{fig:rdd}.
This figure is based on data from \citet{Wuepper2023a}. 
Their starting point is to quantify for each of many years how much countries
matter for local crop yields. 
Here, we only focus on two countries:
Vietnam and Cambodia. 
The border can be divided into small segments (panel~a) 
and crop yields can be quantified in high resolution from satellite imagery (panel~b) \citep{wuepper2025}.
When computing local averages of crop yields at equal distances from the border and plotting these as a
function of border distance, a striking pattern emerges: Whereas crop
yield is distributed rather smoothly on either side of the border, there
is a stark jump right at the border (panel~c), which cannot be explained by
potential confounders such as rainfall or sunshine because these do
\emph{not} jump at the border:
It is the countries as political
constructs that make the fields in Vietnam more productive than those in Cambodia \citep{Wuepper2023a}.
The most important assumption
here is that no potential confounding factors also show a
discontinuity right at the border. For example, if this border was located
right on top of a natural barrier such as a major mountain range, the
sudden change in agricultural conditions could also explain a jump in
crop yields. This can be tested, e.g., by replacing the outcome
variable, in this case crop yields, with elevation, rainfall, temperature, or
sunshine, which would reveal whether these are also discontinuously distributed. 
\citet{Wuepper2023a} analyse first the role of the institutions of these countries in differences in crop yields and, secondly, how much agricultural technology (mechanisation and irrigation) is the channel. 
For these further analyses, they move on to panel data, as discussed in the following paragraph.

An increasingly popular research design is the
Difference-in-Discontinuity Design (DiDD), which is a combination of RDD
and Difference-in-Differences. It is set-up like a standard
DID Design with the only difference being that it
focusses on the change in a discontinuity from before to after treatment.
This built-in extra step improves the chance of a valid parallel trends
assumption because the estimated discontinuity already helps to avoid
confounding factors as discussed above. 
In the best-case scenario, 
a researcher finds a situation in which the threshold is newly created at
some point in time (e.g., an existing state is split into two), which means that demonstrating that there was no discontinuity prior to treatment is
straightforward, and afterwards the discontinuity shows the causal treatment
effect \citep{Garg2021}. Alternatively, in the study by \citet{Wuepper2023a}, 
the leveraged country borders do not change, but they show
that the discontinuities in crop yields are stable before treatment and
change in response to countries' institutional changes.

Finally, Regression Discontinuity in Time (RDiT) tackles endogeneity
by examining a narrow time window around the implementation of a policy, 
where time is used as the running variable 
and the treatment date acts as the threshold.%
\footnote{
RDiT is related to Interrupted Time Series (ITS), 
which is another method that also leverages temporal variation.
However, RDiT requires discontinuity at the cut-off,
bandwidth selection, and strong RDD assumptions
that can be empirically tested
\citep[e.g., via the density test suggested by][]{McCrary2008}.
While ITS can identify changes in a trend without these requirements, it typically requires longer time series and lacks formal tests for violations of key identifying assumptions, such as manipulation or anticipation.
}
This approach assumes that unobserved factors remain similar within the window,
which allows pre-treatment observations to be used as a comparison for post-treatment observations.
RDiT utilises flexible polynomial time trends
and has been recently used in studies involving sin taxes, sugar and fat taxes, air quality, fisheries, and food safety \citep{Hausman2018,Bovay2025}. 
The growing availability of high-frequency data further enhances its utility for researchers
evaluating national agricultural and environmental policies and interventions.

For the research designs discussed above, simple procedures can be followed, which include performing various tests and analytics in a chronological order, 
which allows readers to easily follow and judge the credibility of the analysis \citep{Wuepper2023}. 
This is aided by the implementation of this method in off-the-shelf packages, especially the Python, R, and Stata packages provided by \citet{Calonico2015} and \citet{Calonico2017}.%
\footnote{
All available at:
\url{https://rdpackages.github.io/rdrobust/}
}
The two main assumptions of RDD are exogenous thresholds and no endogenous
sorting. 
The simplest way of examining the assumption of no endogenous sorting is to look for bunching near the threshold \citep{McCrary2008}.
The simple logic is that if there is a striking dip in observations on
one side of the threshold, and these ``missing'' observations all bunch together
on the other side of the threshold, it is likely that it is the result
of optimising behaviour (e.g., if a regulation that only
applies to farms above 5~hectares was introduced, farmers who initially had 5.2
hectares quickly got rid of 0.3 hectares).

There are two important technical aspects to consider, plus a warning.
First, for any kind of discontinuity analysis, one needs to restrict the dataset to observations within an ``optimal'' bandwidth near the cut-off \citep{cattaneo2022regression}.
This can have an important impact on the estimates as it involves a variance-bias trade-off.
The cleanest comparison is possible just next to the threshold (assuming the absence of spillovers here). 
However, using only observations that are directly at the threshold will make the sample size small and specific; keeping only one observation on each side of the threshold would even make it impossible to fit a regression line.
Thus, in order to obtain precise and meaningful estimates, one must allow for some maximum distance to the border,
while still having two sides that are sufficiently comparable to each other.
Over the years, various algorithms have been developed that aim to find the statistically optimal bandwidth \citep{Wuepper2023}.
It is generally a good idea to demonstrate the sensitivity of one's findings to small (or large) deviations from the chosen bandwidth.
For example, if the running variable is farm size and the optimal bandwidth (e.g., according to the Mean Squared Error) is 30~hectares,
it is good to additionally report the findings for a bandwidth of 25 and 35~hectares.
Second, in addition to choosing the optimal bandwidth, one must decide how to fit the regression to the observations.
The simplest approach is to use a linear regression with a dummy variable identifying the threshold and then two variables reflecting the continuous running variable, separately on each side of the threshold.
A more sophisticated way to do it is to use local polynomial functions \citep{cattaneo2022regression}.
These can be based on a continuity assumption, as discussed above,
i.e., a smooth distribution of potential outcomes across the threshold,
or a local randomisation assumption similar to common experimental set-ups,
i.e., potential outcomes are statistically the same on either side of the threshold
\citep{cattaneo2022regression, Wuepper2023}.
A limitation of the local polynomial approach is its relative complexity and computational demand compared to a linear regression framework.
For example, in the local polynomial framework,
it is not straightforward how to handle panel data with fixed effects. 
Furthermore, with very large datasets, such as those that are becoming more common now with the growing availability of high-resolution satellite data \citep{wuepper2025}, 
the simpler linear regression approach is clearly faster than the local polynomial approach.

A question to reflect upon is how generalisable the very locally identified effects of a given discontinuity analysis are. 
Sometimes, a threshold might be found at a comparably extreme end of a range. 
For example, the average farm size in a country might be 100~hectares and the threshold used for identification is five hectares. 
Hence, it needs to be discussed how relevant the results based on farms close to the threshold are for the average farm.

When using discontinuity-based methods, we suggest doing the following
(in addition to following the general suggestions
that we provide in Section~\ref{sec:conclusion}):
\begin{itemize}
\item Visually assess the discontinuity (or the change in discontinuity) and the data distributions around the discontinuity.
\item Conduct placebo tests to probe the exogeneity of the threshold \citep[see, e.g.,][]{Wuepper2023}.
\item Use alternative algorithms to compute the optimal statistical bandwidth for robustness checks.
\item Test for endogenous sorting across the threshold \citep{McCrary2008}.
\item In a discrete running variable with mass points, consider local randomisation or redefining the running variable by aggregating observations at the mass points to handle the discreteness \citep{cattaneo2022regression} . 

\end{itemize}

\section{General Suggestions and Conclusions}
\label{sec:conclusion}

We do not recommend one particular method over
another as whichever method is suitable is case-dependent. Therefore, our aim is to provide
clear guidelines that should be followed when applying these methods.

In addition to the method-specific guidelines provided in previous sections of this paper,
we suggest doing the following irrespective of the chosen method:
\begin{itemize}
\item Before pursuing causal inference, it is important to determine whether the question at hand concerns the ``effect of a cause'' or the ``cause of an effect.'' Plausible and policy-relevant causal inference can typically be made only in relation to the ``effect of a cause,'' not the ``cause of an effect.'' The latter is only meaningful to the extent that it helps identify which cause to study when estimating the ``effect of a cause'' \citep{Gelman2013}. 
\item Start from the theoretical understanding of the problem (e.g., based on a DAG) to define an identification strategy and clearly discuss under what assumptions the quantity of interest is identified, any potential explanations for violating the assumptions and their consequences for identification.
\item Carefully consider the assumptions of various estimation approaches. Consider the extent to which these assumptions fit the theoretically motivated identification strategy.
\item Clearly point out the added value of the chosen method compared to simpler approaches such as OLS.
Unless a relevant added value can be clearly demonstrated, a simpler method may be preferable.
\item Discuss the plausibility of the ``Stable Unit Treatment Value Assumption'' (SUTVA) in your specific empirical analysis.
Under this assumption, the potential outcomes of each observation only depend on the treatment of this observation and not on the treatment of other observations.
All methods discussed in previous sections require this assumption unless spillovers between observations are explicitly and appropriately accounted for in the empirical analysis.
\item Simulate artificial data sets with known properties before using actual data to perform an empirical analysis.
These properties may include the functional form of the analysed relationship,
the magnitude of the treatment effect and its heterogeneity between observations, 
correlations between observed variables and between observed and unobserved variables, 
potential endogeneity issues, 
validity of the exclusion restriction and IV strength (in the case of an IV-based method), 
the degree of serial correlation of observed and unobserved variables (in the case of panel data and/or the use of lagged variables), 
deviations from independently and identically distributed (iid) error terms (e.g., heteroscedasticity, clustering),
and other assumptions. 
Use these data sets to test the estimation approach (as well as the code used to implement it). 
Test under which conditions the estimation approach succeeds in recovering the effects used to create the artificial data. 
Using artificial data to test the code/inference is a integral part of the data-generating-process centric workflow proposed in \citet{Storm2024}.
\item Use multiple approaches and critically discuss what can be learnt from the results of different methods as, in most cases, there may not be a single best estimation approach as each approach has its advantages/drawbacks.
Recent textbooks on causal inference, such as \citet{Cunningham2021} and \citet{HuntingtonKlein2025},
provide more detailed information about and code examples for several of the methods mentioned in this paper,
which are helpful sources of information for robustness checks and sensitivity analyses.
\end{itemize}

Even if these guidelines are followed, when investigating causal effects with observational data,
there is always uncertainty about whether all the required assumptions are completely fulfilled. 
Therefore, one should be very careful when using causal language such as
``the effect of A on B'', 
``the impact of A on B'', 
``A affects B'', 
``A reduces B'', 
``A increases B'', 
``A leads to a change in B'', etc.
As a precaution, one could use statements about associations such as
``A is positively related to B'',
``A is negatively related to B'',
``A is associated with B'',
``A is conditionally associated with B'', etc. 
In any case, it is important to use consistent language throughout the entire paper.%
\footnote{
One minor exception to this rule would be to write that a study ``aims to estimate the effect of A on B'', to explain why the estimates may not indicate causal effects, and to interpret all estimates as conditional associations \citep[as done in, e.g., ][]{Aihounton2024}.
}
If causal statements are made, it is crucial to clearly point out that these statements are conditional on the appropriateness of the identifying assumptions, the model specification implemented, and the data used for the estimation.

\section*{Acknowledgements}

The authors would like to thank Marc F.\ Bellemare, Irene Villalba, and two anonymous reviewers for their feedback on earlier versions of this manuscript and Jonathan Roth and Pedro Sant'Anna for their valuable feedback on DID methods.
Arne Henningsen extends his gratitude to Marc F.\ Bellemare and Fabio G.\ Santeramo
for their invaluable discussions and inspiration, 
especially during the joint preparation and teaching of PhD courses
on the topic of this paper.
Arne Henningsen is also grateful for inspiring discussions with other members of the research project
``CliFT -- Climate-Smart Futures in Rural Tanzania''
and to the Ministry of Foreign Affairs of Denmark for financially supporting this project
(Grant: 23-04-KU). 
David Wuepper and Hugo Storm acknowledge the financial support 
provided by the German Research Foundation (DFG) Cluster of Excellence
``PhenoRob'' (EXC 2070 Grant No.\ 390732324). 
Guy Low and Tobias Dalhaus acknowledge the financial support 
provided by the European Union's Horizon 2020 research and innovation programme 
(grant agreement No.\ 862357 (MIXED)); 
this work only reflects the authors' views;
the Commission is not responsible for any use of the information contained in this article.
Dagim Belay acknowledges the financial support 
provided by Innovation Fund Denmark (grant No.\ JPIAMR2021-182 SEFASI) 
under the umbrella of the Joint Programming Initiative on Antimicrobial Resistance (JPIAMR) and the Independent Research Fund Denmark (DFF)
(grant No.\ 2029-00021B).
Of course, the authors take full responsibility for any remaining errors.

\bibliographystyle{apalike}
\bibliography{references}

\clearpage
\appendix
\section*{Appendix}

\section{Extended IV Methods}
\label{sec:IV-extended-methods}

While the discussions in Section~\ref{sec:IV} refer to IV and 2SLS regression, 
they are largely transferable to other methods
that rely on IVs
such 3-stage least squares (3SLS) regression, 
extended IV methods for binary endogenous regressors 
(\citealp[p.~142--144]{Angrist2009}; \citealp[][p.~937--942]{Wooldridge2010}; \citealp{Wooldridge2015}), 
and more recent estimators that
are particularly suited to handling binary and ordinal endogenous
variables such as the extended regression IV approaches in Stata, which
estimate the parameters using maximum likelihood (see \citealp{Jafari2023}, 
for an example and \citealp{StataPress2023}, p.~183, for a technical description). 
These discussions are also largely transferable to
estimators that are based on distributional assumptions of error terms
as suggested by \citet{Heckman1976} such as the endogenous treatment effect
model and the endogenous switching regression model. 
These models can be estimated with a two-stage approach that uses an inverse Mills ratio as
additional regressor in the second-stage regression or with a one-step
maximum likelihood estimation. In fact, these models can be estimated
without instrumental variables (or exclusion restrictions) but in this
case, the identification of the estimated parameters hinges solely on
the distributional assumptions, e.g., a bivariate normal distribution of
the two error terms. As it is very unlikely that the distributional
assumptions will be fulfilled exactly in a real-world application, using
these estimators without instrumental variables would very likely result in
unreliable estimates. As strong instrumental variables render the
distributional assumptions less relevant, it is imperative to use strong
instrumental variables when using these estimators. Thus, at least one variable that strongly affects the selection outcome (i.e., whether an
observation is treated in an endogenous treatment effect model or
whether an observation is in the first or second outcome regime of an
endogenous switching regression model) but does not affect the
dependent variable of the outcome equation and is not related to the
error term(s) of the outcome equation(s) is needed \citep[see, e.g.,][for an example]{Auci2021}. 
These variables are frequently called
instrumental variables because they basically need to fulfil the same
criteria as instrumental variables in the regression methods discussed
in the beginning of this section. 
Hence, the validity of the exclusion
restrictions must be investigated and critically discussed in similar
ways to the validity of instrumental variables in the regression methods
discussed in the beginning of this section.

A straight-forward extension of a 2SLS estimation to non-linear
regression models would be to regress each endogenous explanatory
variable on the exogenous explanatory variables and the instrumental
variables (using linear or non-linear regression) and to obtain the
predicted values of the endogenous explanatory variables. One can then
estimate the non-linear regression model with the endogenous explanatory
variables replaced by the predicted values obtained in the first stage.
However, caution is advised here to avoid falling into what \citet{Angrist2009} 
refer to as the ``forbidden regression'' trap and
directly applying the 2SLS argument to a non-linear case, for example,
using the predicted values from a probit first-stage in the second
stage. Another mistake that must be avoided in this context is, when
dealing with both a linear and quadratic form of the endogenous
variable, simply using the square of the predicted values from the
first-stage instead of estimating two separate first-stage regressions
\citep{Angrist2009}. 

In the case of non-linear least-squares
regression, the non-linear two-stage least squares (N2SLS) estimator has
similar properties to the 2SLS estimator \citep{Amemiya1974}. However, in
many other non-linear regression models (e.g., logit, probit, count-data
models), this approach, which is sometimes called two-stage predictor
substitution (2SPS), results in inconsistent estimates \citep[e.g.,][]{Terza2008}. 
An alternative to this approach is a slightly different
procedure: The first stage is identical to the first-stage regression of
2SLS, N2SLS and 2SPS estimators, but in the second stage, the residuals
that were obtained in the first stage are added as additional regressors
(while the endogenous explanatory variables are used as regressors).
This approach is called Two-Stage Residual Inclusion (2SRI) in
biostatistics and health economics \citep[e.g.,][]{Terza2008}, while it
is called the control-function (CF) approach in the econometrics literature
\citep[e.g.,][]{Wooldridge2015}. In the case of linear regression models, this
approach provides the same estimates as a 2SLS estimation, while the
consistency of this approach has been demonstrated for many non-linear
estimators. Hence, it is frequently used to address the endogeneity of
regressors in non-linear regression models such as double hurdle models
\citep[e.g.,][]{Rao2013,Sellare2020b} or fractional logit
models \citep[e.g.,][]{Wuepper2020a}. 
As the identifying assumptions for the
2SRI / CF approach are similar to those of IV and 2SLS estimations, 
the identification strategy should be based on the same
evaluation criteria as for other estimations with IVs.

A further regression framework that can be used in an instrumental-variable setting is
the Generalised Method of Moments (GMM), which identifies the regression coefficients by assuming moment conditions in
the population and then imposing these moment conditions in the sample.
The number of assumed moment conditions must be equal to or larger than
the number of regression coefficients to be estimated. Given that a
myriad of different moment conditions can be assumed, the GMM framework
is very flexible and many well-known estimators such as OLS regression
and 2SLS regression are special cases. If a GMM approach is used to
estimate causal effects, the appropriateness of the assumed moment
conditions must be thoroughly and critically discussed. If a GMM
estimation uses instrumental variables, the validity of these IVs should
be discussed in a similar way as for other methods that use IVs. If we have more moment conditions available than we have regression
coefficients, a Sargan-Hansen test (also known as Sargan's \emph{J} test
or Hansen's \emph{J} test) can be used to empirically assess the
validity of the moment conditions. 

In the case of panel data, the GMM
framework can address the endogeneity of explanatory variables even without
external instruments by using the lagged values of some variables as
`internal' instruments. The ``Difference GMM'' estimator suggested by
\citet{Arellano1991} and the ``System GMM'' estimator suggested by
\citet{Arellano1995} and \citet{Blundell1998} are frequently
used GMM estimators that use internal instruments. The moment conditions
assumed by these types of estimators can be complex. Similar to using
lagged values of endogenous regressors as IVs in 2SLS estimations \citep[see Section~\ref{sec:IV-special-types} below and][]{Wang2020}, 
these types of estimators usually
require restrictive assumptions about unobserved
factors, which may be unrealistic in most empirical applications.

Even if one uses a valid
(i.e., exogenous and highly relevant) instrument, 
IV regression can result in substantially biased estimates
if parametric assumptions,
e.g., about the functional form,
are not fulfilled \citep{Okui2012}.
Hence, it might be worthwhile
to consider using non-parametric IV regression methods.
\citet{Chernozhukov2018} show that Double Machine Learning 
(see Section~\ref{sec:selection-on-observables}) 
can also be applied to an IV setting, which means the
linearity assumption of 2SLS regression can be relaxed. 
Their approach allows both the outcome equation and the treatment equation
to be unknown nonlinear equations
that can be approximated by any flexible machine learning algorithm. 
However, it still requires assuming either homogeneity of treatment or homogeneity
of treatment assignment. Under these conditions, the approach provides a
consistent estimate of an average treatment effect (ATE). Going further,
multiple approaches also relax the homogeneity assumptions and allow
the estimation of treatment effects that vary depending on the observed
characteristics. \citet{Hartford2017} have developed an approach called
DeepIV, which uses deep neural networks in both the outcome and the treatment model. 
\citet{Athey2019} have developed Generalised Random Forests (RFs) as a
nonparametric estimator that can be used to estimate any quantity
identified by a set of (local) moment conditions. They demonstrate that this
approach can be used to estimate treatment effects under the
unconfoundedness assumption (leading to an approach called Causal
Forests, see Section~\ref{sec:selection-on-observables}) 
but also in an IV setting. Generalised RFs can
basically be understood as a more flexible alternative to GMM
estimation methods. Importantly, Generalised RFs are able to learn
treatment heterogeneity in a data-driven manner. Additionally, it is
possible to obtain asymptotic uncertainty intervals for the estimated
treatment effect, allowing the user to assess uncertainty in the estimates and
perform hypothesis testing. While DeepIV and Generalised RFs are
specifically designed around deep neural networks and RFs, respectively,
\citet{Syrgkanis2019} provide a generalised framework (Orthogonal IV)
for nonparametric IV estimations that allows the use of any machine learning
approach in the outcome and treatment model. They also develop methods
that allow the projection of treatment heterogeneity to a simpler (potentially
linear) lower dimensional space. This means asymptotic
confidence intervals can be derived and machine learning interpretability
methods (e.g., SHAP values) can be used to illustrate and inspect treatment
heterogeneity.

Another relatively specialised case of machine learning in the
context of IV estimation is to deal with a situation in which there is a
large number of potential instruments (potentially larger than the
number of observations). \citet{Belloni2012} demonstrate that simple machine
learning methods such as LASSO can be used to select instruments under
the assumption that the treatment assignment can be sufficiently predicted
by a small subset of all the available instruments. However, in
empirical settings, we very rarely face the (luxury) problem of having too many IVs.

Most of the machine-learning approaches 
that are relevant for applied economists
(Double Machine Learning, DeepIV, Causal Forest, Generalised RFs for IV,
Orthogonal IV) are available in the Python package EconML
(\url{https://econml.azurewebsites.net/index.html}), which provides a unified API for all these approaches 
and represents a relatively simple application for applied researchers.

\section{Special Types of Instruments}
\label{sec:IV-special-types}

This section discusses some special types of instrumental variables
that are frequently used in agricultural and applied economics.
One of these special types of instrumental variables is the so-called spatial instrumental variable or leave-one-out
instrumental variable \citep[e.g.,][]{Mason2013,Krishnan2014,Smale2014,Magnan2015,Wuepper2018,Sellare2020a,TabeOjong2022,Aihounton2024}. 
In this case, an endogenous explanatory (treatment) variable is
instrumented by the average or proportion within a peer group leaving
out the respective observation. For example, a farmer's adoption of a
technology is instrumented by the proportion of farmers in the village
who adopted this technology leaving out the respective farmer. However,
while this type of instrumental variable is usually highly relevant, its
exogeneity requires strict assumptions that are not fulfilled in many
empirical applications \citep{Angrist2014,Betz2018,McKenzie2018}. 
In some empirical analyses, it may be reasonable to use such a
spatial instrumental variable or a variant thereof, potentially combined
with other tools, but authors must provide clear reasoning as to why this
identification strategy is valid in their study \citep[e.g.,][]{Maggio2022}. 

Closely related to spatial instruments are Hausman-type
instruments, which are frequently used in food product demand analyses to
account for the endogeneity of product prices \citep[see, e.g.,][]{Nevo2001}.
The idea is that the price of a product in other regions can be used as
instrument since the same product has similar marginal costs across
regions but different demand shifters \citep{Hausman1996,Nevo2000,Hirsch2018}. 
However, this assumption may be violated in the case of a
nationwide shock in demand, for example, if a nationwide advertising
campaign that influences the demand of a product across regional
borders is launched \citep{Nevo2000,Nevo2001}.

Similar to using lagged values of explanatory variables to address
endogeneity in an identification-on-observables identification strategy
(see Section~\ref{sec:selection-on-observables}),
lagged values can also be used as instrumental variables; an
identification strategy that is popular among applied economists.
However, \citet{Wang2020} show that IVs of this type require
specific assumptions. For instance, even if the exclusion restriction
is fulfilled, the estimates are biased (although
consistent), and the likelihood of making Type-1 errors is high if there
is first-order autocorrelation in unobserved factors because this leads
to a correlation between the lagged IV and the error term \citep{Wang2020}. 
As this cannot be ruled out in most empirical
applications, \citet{Wang2020} conclude that using lagged
values of endogenous explanatory variables as instrumental variables
``is unlikely to lead to credible estimates.''

Shift-share instruments, also known as Bartik-type instruments \citep{Bartik1991,Borusyak2025}, 
can be used in cases where panel data is available and the intensity of a unit's treatment is affected by an initial share that affects the exposure to a trend.
Either the trend or the share need to be exogenous for this approach to be valid. 
Then, the interaction of the shift and the share provides an instrumental variable, conditional on the standard IV assumptions being valid. 
For example, when analysing the effect of a regional
subsidy on farm performance, 
a shift-share instrument can be constructed
based on the idea that the nationwide values of subsidies ``shift'' the
regional (endogenous) subsidies according to a predetermined
out-of-sample economic state of the region (share) \citep[see, e.g.,][for an example]{Zou2024}. 
More precisely, in this case, the Bartik IV is
the product of a variable 
that captures the national subsidy level and a
variable with information on the initial state of the regional economy, e.g.,
one year before the start of the sample period that is used in the analysis.
This reflects the exogenous variation in regional subsidies which is uncorrelated with the regional-level error term, 
which means that it may serve as a valid IV \citep{Bartik1991,Breuer2022,Zou2024}. 
It is important to note that for shift-share instruments, 
valid identification can be achieved
when either the shift component
or the share component of the IV is exogenous. 
For additional guidance, we refer to \citet{Borusyak2025}. 
Another illustrative example of a shift-share IV analysis is the paper of \citet{gollin2021} 
who estimate the impact of the Green Revolution with a shift-share~IV.

\end{document}